\documentclass{paper}
\usepackage{graphicx}
\usepackage{amssymb,amsmath,amsfonts}

\def\be{\begin{eqnarray}}
\def\ee{\end{eqnarray}}

\begin{document}

\thispagestyle{empty}

\hfill ITEP/TH-23/08

\bigskip

\centerline{\Large{Analogue of the identity Log Det = Trace Log for resultants
}}

\bigskip

\centerline{\emph{A.Morozov and Sh.Shakirov}}

\bigskip

\centerline{ITEP, Moscow, Russia}

\centerline{MIPT, Dolgoprudny, Russia}

\bigskip

\centerline{ABSTRACT}

\bigskip

{\footnotesize
Like evaluation of Gaussian integrals is based on determinants, exact (non-perturbative) evaluation of non-Gaussian integrals is related to algebraic quantities called resultants. Resultant $R_{r_1, \ldots, r_n}$ defines a condition of solvability for a system of $n$ homogeneous polynomials of degrees $r_1, \ldots, r_n$ in $n$ variables, just in the same way as determinant does for a system of linear equations. Because of this, resultants are important special functions of upcoming non-linear physics and begin to play a role in various topics related to string theory. Unfortunately, there is a lack of convenient formulas for resultants when the number of variables is large. To cure this problem, we generalize the well-known identity Log Det = Trace Log from determinants to resultants. The generalized identity allows to obtain explicit polynomial formulas for multidimensional resultants: for any number of variables, resultant is given by a Schur polynomial. We also give several integral representations for resultants, as well as a sum-over-paths representation.}

\tableofcontents

\section{Introduction}

\emph{Resultants} and closely related \emph{discriminants} are the central objects in emerging field of \emph{non-linear algebra} \cite{GKZ,NOLINAL}, which are generalizations of \emph{determinants} from linear to non-linear systems of algebraic equations. \linebreak While the vanishing of determinant is a condition of solvability for a linear system, the vanishing of resultant is a condition of solvability for a non-linear system. For example, a linear system
$$ \left\{ \begin{array}{c}
f_{11} x_1 + f_{12} x_2 = 0 \\
\noalign{\medskip}f_{21} x_1 + f_{22} x_2 = 0 \\
\end{array} \right. $$
is solvable (has a non-trivial solution $x_1, x_2$), only if its determinant vanishes:
\begin{center}
{\fontsize{9pt}{0pt}
$R_{1,1}(f) = f_{1 1} f_{2 2} - f_{1 2} f_{2 1} = 0$}
\end{center}
In complete analogy, a quadratic system
\[ \left\{ \begin{array}{c}
f_{111} x_1^2 + f_{112} x_1 x_2 + f_{122} x_2^2 = 0 \\
\noalign{\medskip}f_{211} x_1^2 + f_{212} x_1 x_2 + f_{222} x_2^2 = 0 \\
\end{array} \right. \]
is solvable only if its resultant vanishes:
\begin{center}
{\fontsize{9pt}{0pt}
$R_{2,2}(f) = f_{1 1 1}^2 f_{2 2 2}^2 - f_{1 1 2} f_{1 2 2} f_{2 1 2} f_{2 1 1} + f_{1 1 2}^2 f_{2 1 1} f_{2 2 2} - 2 f_{1 1 1} f_{1 2 2} f_{2 1 1} f_{2 2 2}+ f_{1 1 1} f_{1 2 2} f_{2 1 2}^2 - f_{1 1 1} f_{1 1 2} f_{2 1 2} f_{2 2 2} + f_{1 2 2}^2 f_{2 1 1}^2 = 0$}
\end{center}

Resultants and determinants have a lot in common. However, resultants are worse understood, than determinants. Historically, this is because non-linear studies were largely abandoned in favour of linear methods. Matrices, linear operators, determinants and traces became widely accepted tools, much more popular than their non-linear analogues. Linear algebra has found countless applications in physics and in QFT, where it is used to describe Gaussian systems -- such as free particles, non-interacting fields, etc. Much more interesting non-Gaussian systems are usually described perturbatively, i.e. as approximately Gaussian.

\smallskip

The situation is changing in our days, when non-perturbative treatment of non-linear physics attracts more and more attention. Resultants and discriminants start to play an increasing role in physical applications (see \cite{Du,Va,MN,im8} for some recent examples), which, however, remain restricted by the lack of developed theory for these very important quantities. Very little is known about resultants and discriminants for many variables. Even explicit polynomial formulas for resultants, analogous to classical formulas for determinants
\begin{equation}
\begin{array}{cc}
R_{1}(f) = det_{1 \times 1}(f) = f_{11} \\
\\
R_{1,1}(f) = det_{2 \times 2}(f) = f_{1 1} f_{2 2}-f_{1 2} f_{2 1} \\
\\
R_{1,1,1}(f) = det_{3 \times 3}(f) = f_{1 1} f_{2 2} f_{3 3}-f_{1 1} f_{2 3} f_{3 2}-f_{2 1} f_{1 2} f_{3 3}+f_{2 1} f_{1 3} f_{3 2}+f_{3 1} f_{1 2} f_{2 3}-f_{3 1} f_{1 3} f_{2 2} \\
\end{array}
\label{Expl}
\end{equation}
remain an open issue. Such polynomial expressions are known in low-dimensional cases (see chapter 13 of \cite{GKZ} and chapter 5 of \cite{NOLINAL} for more details), but in higher dimensions they are not found yet. In this paper, we hope to give a solution of this problem.

\smallskip

In higher dimensions, study of resultants is complicated by the extreme length of these quantities. If written in the form (\ref{Expl}), the simplest multidimensional resultant $R_{2,2,2}$ contains $\sim 22000$ terms and takes about $50$ standart $A4$ pages! However, by itself this large size is not a real problem. As we know, explicit expressions for determinants are also rather lengthy: determinant $$ R_{\underbrace{1,1,\ldots,1}_{n}} = det_{n \times n} $$ contains $n!$ terms, much less than $R_{r_1,\ldots,r_n}$ with $r_i > 1$, but still a lot. What one really needs to know are the \emph{properties} of resultants: the structure, which stands behind the numerous terms. Like in the case of determinants, this knowledge will provide effective ways to work with these objects.

\smallskip

An approach, suggested in \cite{NOLINAL}, is to study resultant theory as a natural generalization of determinant theory.
Indeed, all well-known properties of determinants have analogues for resultants. Explicit description of these analogues, originated in \cite{NOLINAL}, is the subject of this, related \cite{NOLINAL,Dol,Sha} and forthcoming \cite{Ano,LogDet} papers. We are especially interested in constructive formulas. Linear algebra has worked out many such formulas for determinants: expansion by minors, expression of determinants through traces via $\log$ Det = Trace $\log$, invariant $\epsilon$-tensor formulation, eigenvalue representation, etc. It is natural to look for their direct counterparts for resultants.

\smallskip

Following this line of research, in this paper we generalize to resultants the relation log Det = Trace log. This famous identity allows to express a determinant through a much simpler operation -- trace:
\be
 {\rm log \ det}(I - f) = {\rm tr \ log}(I - f) = - \sum\limits_{k = 1}^{\infty} \dfrac{{\rm tr }f^k}{k} = - {\rm tr }f - \dfrac{1}{2} {\rm tr }f^2 - \dfrac{1}{3} {\rm tr }f^3 + \ldots
\label{LogDet}
\ee
where $f: {\mathbb C}^n \rightarrow {\mathbb C}^n$ is a linear map, defined by $n$ homogeneous linear polynomials
$$
f: \left( \begin{array}{c} x_1 \\ x_2 \\ \ldots \\ x_n \end{array} \right) \mapsto \left( \begin{array}{c} f_{11} x_1 + f_{12} x_2 + \ldots + f_{1n} x_n \\ f_{21} x_1 + f_{22} x_2 + \ldots + f_{2n} x_n \\ \ldots \\ f_{n1} x_1 + f_{n2} x_2 + \ldots + f_{nn} x_n \end{array} \right)
$$
$I$ is the unit map, or simply unity
$$
I: \left( \begin{array}{c} x_1 \\ x_2 \\ \ldots \\ x_n \end{array} \right) \mapsto \left( \begin{array}{c} x_1 \\ x_2 \\ \ldots \\ x_n \end{array} \right)
$$
and trace is the sum of diagonal elements: $${\rm tr }f = f_{11} + \ldots + f_{nn}$$ The shortest way to understand (\ref{LogDet}) is to express both sides in terms of eigenvalues, when $f$ is brought to diagonal (or Jordanian) form. Identity (\ref{LogDet}) is included in all linear-algebra textbooks and has numerous applications in mathematics and physics. For us, it is especially important that (\ref{LogDet}) produces explicit polynomial formulas for determinants. If exponentiated, it turns into
$$ {\rm det}(I - f) = {\rm exp } \left( - \sum\limits_{k = 1}^{\infty} \dfrac{{\rm tr }f^k}{k} \right) = 1 - \underbrace{ \left(tr f\right) }_{det \ 1 \times 1 } + \underbrace{ \dfrac{1}{2} \left((tr f)^2 - tr f^2 \right)}_{det \ 2 \times 2 } - \underbrace{\dfrac{1}{6} \left(2 tr f^3 - 3(tr f)(tr f^2) + (tr f)^3 \right)}_{det \ 3 \times 3 } + \ldots$$
which is an explicit series expansion for ${\rm det}(I - f)$ in powers of $f$. We can recall that in dimension $n$ determinant has degree $n$ in $f$. Consequently, it can be found by collecting terms of that degree in the r.h.s:
\[\begin{array}{cc}
det_{1 \times 1}(f) = tr f \\
\\
det_{2 \times 2}(f) = \dfrac{1}{2} \left((tr f)^2 - tr f^2 \right) \\
\\
det_{3 \times 3}(f) = \dfrac{1}{6} \left(2 tr f^3 - 3(tr f)(tr f^2) + (tr f)^3 \right) \\
\end{array}\]
and so on. As we see, determinants are expressed as certain polynomials of traces. These polynomials are well known as Schur polynomials. They are usually denoted as $P_k$ and defined by
$$\exp \left( \sum\limits_{k = 1}^{\infty} t_k z^k \right) = \sum\limits_{k = 1}^{\infty} P_k \left\{ t_i \right\} \ z^k$$
where the figure brackets denote the set of arguments: $ P_k \left\{ t_i \right\} = P_k \left( t_1, t_2, \ldots \right) $. First Schur polynomials are
\[
\begin{array}{ccc}
P_0\left\{ t_i \right\} = 1 \\
\\
P_1\left\{ t_i \right\} = t_1 \\
\\
P_2\left\{ t_i \right\} = t_{2} + t_{1}^2/2 \\
\\
P_3\left\{ t_i \right\} = t_{3} + t_{1} t_{2} + t_{1}^3/6
\end{array}
\]
and so on, and the explicit expression for $P_k\left\{ t_i \right\}$ is
\be
P_{k}\left\{ t_i \right\} = \sum\limits_{m = 1}^{k} \ \sum\limits_{v_1 + v_2 + \ldots + v_m = k} \ \dfrac{t_{v_1} t_{v_2} \ldots t_{v_m}}{m!}
\label{Schur}
\ee
where the sum is taken over all \emph{ordered partitions} of $k$ into $m$ positive parts. An ordered partition is a way of writing an integer as a sum of integers where the order of the items is significant. In other words, the sum is taken over all $m$-tuples of positive integers $(v_1,v_2,\ldots,v_m)$ such that $v_1 + v_2 + \ldots + v_m = k$. In our case, $t_k = - {\rm tr }f^k/k$ and determinant is a $n$-th Schur polynomial of these quantities (taken with a sign):
$$det_{n \times n}(f) = (-1)^n P_n \left\{ - \dfrac{tr f^k }{k} \right\}$$
From (\ref{Schur}) we obtain
\be
\boxed{
det_{n \times n}(f) = \sum\limits_{m = 1}^{n} \ \dfrac{(-1)^{m + n}}{m!} \sum\limits_{k_1 + k_2 + \ldots + k_m = n} \dfrac{{\rm tr }\ f^{k_1} \ {\rm tr } \ f^{k_2} \ \ldots \ {\rm tr }\ f^{k_m}}{k_1 k_2 \ldots k_m}}
\label{MainLinear}
\ee

Of course, the fact that determinant is given by a Schur polynomial of traces of powers, is very well known. What is interesting, this fact has a direct generalization to resultants for arbitrary number of variables -- and this provides explicit polynomial formulas for resultants. Just like determinants apply to linear maps, resultants $R_{r_1, \ldots, r_n}(f)$ apply to polynomial maps $f: \mathbb C^{n} \rightarrow \mathbb C^{n}$
$$f: \left( \begin{array}{c} x_1 \\ x_2 \\ \ldots \\ x_n \end{array} \right) \mapsto \left( \begin{array}{c} f_1(x_1,x_2,\ldots,x_n) \\ f_2(x_1,x_2,\ldots,x_n) \\ \ldots \\ f_n(x_1,x_2,\ldots,x_n) \end{array} \right), \ \ \ \ \ I: \left( \begin{array}{c} x_1 \\ x_2 \\ \ldots \\ x_n \end{array} \right) \mapsto \left( \begin{array}{c} (x_1)^{r_1} \\ (x_2)^{r_2} \\ \ldots \\ (x_n)^{r_n} \end{array} \right) $$
given by $n$ homogeneous polynomials $f_1, f_2, \ldots, f_n$ of arbitrary degrees $r_1, r_2, \ldots, r_n$. Linear maps correspond to the particular case of $(r_1, r_2, \ldots, r_n) = (1,1, \ldots, 1)$ and resultants of linear maps are just determinants. The map $I$, which plays the role of unity, is called a \emph{diagonal map}. The resultant of $I$ is one: $R_{r_1, \ldots, r_n}(I) = 1$, therefore ${\rm log} \ R_{r_1, \ldots, r_n}(I - f)$ posesses a series expansion in powers of $f$:
\be
 \log \ R_{r_1, \ldots, r_n}(I - f) = - \sum\limits_{k = 1}^{\infty} \dfrac{ T_{k} \left( f \right) }{k}
\label{LogRez}
\ee
where $T_{k} \left( f \right)$ is a homogeneous expression of degree $k$ in $f$. This is a very interesting series expansion -- in fact, a direct analogue of (\ref{LogDet}) for resultants. In this paper we study this expansion, using the methods of non-linear algebra. Our main result is an explicit formula

\be
 \log \ R_{r_1, \ldots, r_n}(I - f) = \left. \left[ \ \sum\limits_{k = 0}^{\infty} \dfrac{r_1}{(r_1 k)!} ({\hat f_1})^{k} \ \right] \left[ \ \sum\limits_{k = 0}^{\infty} \dfrac{r_2}{(r_2 k)!} ({\hat f_2})^{k} \ \right] \ldots \left[ \ \sum\limits_{k = 0}^{\infty} \dfrac{r_n}{(r_n k)!} ({\hat f_n})^{k} \ \right] \cdot \log \ \det(I - A) \right|_{A = 0}\\ \nonumber
\label{LogRezEx}
\ee
where $A$ is an auxillary $n \times n$ matrix, and ${\hat f}_1, {\hat f}_2, \ldots, {\hat f}_n$ are differential operators
\[
\begin{array}{ccc}
{\hat f_1} =  f_1 \left( \dfrac{\partial}{\partial A_{11}}, \dfrac{\partial}{\partial A_{12}}, \ldots, \dfrac{\partial}{\partial A_{1n}} \right) \\
\noalign{\medskip} {\hat f_2} =  f_2 \left( \dfrac{\partial}{\partial A_{21}}, \dfrac{\partial}{\partial A_{22}}, \ldots, \dfrac{\partial}{\partial A_{2n}} \right) \\
\noalign{\medskip} \ldots \\
\noalign{\medskip} {\hat f_n} =  f_n \left( \dfrac{\partial}{\partial A_{n1}}, \dfrac{\partial}{\partial A_{n2}}, \ldots, \dfrac{\partial}{\partial A_{nn}} \right) \\
\end{array}
\]
associated with polynomials $f_1, f_2, \ldots, f_n$. Homogeneous components of the expansion are given by
\be
\dfrac{ T_{k}(f)}{k} \ = \ \sum\limits_{k_1 + k_2 + \ldots + k_n = k} \left. \ \dfrac{r_1 ({\hat f_1})^{k_1}}{(r_1 k_1)!} \ldots \dfrac{r_n ({\hat f_n})^{k_n}}{(r_n k_n)!} \ \cdot \dfrac{ tr A^{r_1 k_1 + \ldots + r_n k_n} } {r_1 k_1 + \ldots + r_n k_n } \ \right|_{A = 0}
\label{NTraces}
\ee
Quantities $T_{k}(f)$ are called \emph{traces} for a non-linear map $f$. One can easily check, that for linear polynomials they turn into ordinary traces $tr f^k$, reproducing the conventional relation Log Det = Trace Log. Analogy with linear algebra goes further: as long as the traces are known, resultants are expressed through them:
$$
 R_{r_1, \ldots, r_n}(I - f) = {\rm exp } \left( - \sum\limits_{k = 1}^{\infty} \dfrac{T_{k}(f)}{k} \right) = 1 - T_1 + \dfrac{1}{2} \left(T_1^2 - T_2\right) - \dfrac{1}{6} \left(2 T_3 - 3 T_1 T_2 + T_1^3 \right) + \ldots
$$
Since resultant $R_{r_1,\ldots,r_n}(f)$ is known \cite{GKZ,NOLINAL} to be a homogeneous polynomial in $f$ of degree $$ d = r_1 r_2 \ldots r_n \left( \dfrac{1}{r_1} + \dfrac{1}{r_2} + \ldots + \dfrac{1}{r_n} \right) $$ it can be found by collecting terms of that degree in the r.h.s (just like we did for determinants):
\[
\begin{array}{ccc}
R_{1,1} = \dfrac{1}{2} \left( T_{1}^2 - T_{2} \right) \\
\\
R_{1,1,1} = \dfrac{1}{6} \left( T_{1}^3 - 3 T_{1} T_{2} + 2 T_{3} \right) \\
\\
R_{2,2} = \dfrac{1}{24} \left( T_{1}^4 - 6 T_{1}^2 T_{2} + 8 T_{1} T_{3} + 3 T_{2}^2 - 6 T_{4} \right) \\
\\
R_{1,1,2} = \dfrac{1}{120} \left( T_{1}^5-10 T_{1}^3 T_{2}+20 T_{1}^2 T_{3}+15 T_{1} T_{2}^2-30 T_{1} T_{4}-20 T_{2} T_{3}+24 T_{5} \right)\\
\end{array}
\]
and so on, and the explicit expression for $R_{r_1,\ldots,r_n}(f)$ is
$$R_{r_1,\ldots,r_n}(f) = (-1)^d P_{d} \left\{ - \dfrac{T_k(f) }{k} \right\}$$
In this way, resultants are expressed as Schur polynomials of $t_k = -T_k/k$. From (\ref{Schur}) we obtain
\be
\boxed{
R_{r_1,\ldots,r_n}(f) = \sum\limits_{m = 1}^{d} \ \dfrac{(-1)^{m + d}}{m!} \sum\limits_{k_1 + k_2 + \ldots + k_m = d} \dfrac{T_{k_1}(f) \ T_{k_2}(f) \ \ldots \ T_{k_m}(f)}{k_1 k_2 \ldots k_m}}
\label{Main}
\ee
which is, in fact, an explicit polynomial formula for a resultant $R_{r_1,\ldots,r_n}$ of $n$ homogeneous polynomials in $n$ variables. Traces $T_{k}(f)$ are calculated using the polynomial expression (\ref{NTraces}). It is explicit and can be straightforwardly implemented on any computer system.

\smallskip

Formula (\ref{Main}) is a useful complement and alternative to the Cayley-Sylvester-Bezout approach, which is based on \emph{determinantal} formulas and surveyed in \cite{GKZ,NOLINAL,Ano}. Such formulas express resultants through determinants of various auxillary matrices and, in general, of certain Kozhul complexes. While effective in low-dimensional cases, this approach becomes inpractical in higher dimensions. This is because i) the choice of matrices and complexes in general case is a kind of art and ii) the number and size of auxillary matrices, needed to calculate a resultant, grows too fast. In higher dimensions (\ref{Main}) is expected to be i) less ambigous and ii) more efficient.

\smallskip

The existence of relations like (\ref{LogRezEx}) is also important from pure theoretical point of view. As we know, (\ref{LogDet}) is crucial for matrix-model considerations, and its generalization (\ref{LogRezEx}) could provide a new tool in the study of matrix models \cite{mamo} or -- even more generally -- could help to develop a functional-integral representation of non-linear algebra. The real motivation is inverse: to express non-Gaussian integrals (ordinary or functional) through well-defined algebraic quantities like resultants and discriminants. This very interesting issue will be briefly touched in the section 6 below, but mostly left beyond the scope of this paper.

\section{Definitions and notations}

In this paper, we treat all polynomials as homogeneous. A homogeneous polynomial of degree $r$ in $n$ variables $x_1, x_2, \ldots, x_n$ is a function
$$ f( {\vec x} ) = f \left( x_1, x_2, \ldots, x_n \right) = \sum_{i_1 \leq i_2 \leq \ldots \leq i_r} f_{i_1 i_2 \ldots i_r} \cdot x_{i_1} x_{i_2} \ldots x_{i_r} $$
associated with a completely symmetric tensor $f_{i_1 i_2 \ldots i_r}$ with $r$ indices, taking values from $1$ to $n$. A basic goal of non-linear algebra is to study exact solutions of polynomial equations and systems of equations. A system of $k$ homogeneous polynomials (of arbitrary degrees)
\[
\left\{
\begin{array}{lll}
f_1 \left( x_1, x_2, \ldots, x_n \right) = 0\\
\noalign{\medskip}f_2 \left( x_1, x_2, \ldots, x_n \right) = 0\\
\noalign{\medskip}\ldots \\
\noalign{\medskip}f_k \left( x_1, x_2, \ldots, x_n \right) = 0\\
\end{array}
\right.
\]
in general position has $n - k$ dimensional space of solutions. For example, one equation
$$ f_1 \left( x_1, x_2, \ldots, x_n \right) = 0 $$
defines a surface of dimension $n - 1$ in the $n$-dimensional space $\left( x_1, x_2, \ldots, x_n \right)$, while a pair of equations
\[
\left\{
\begin{array}{lll}
f_1 \left( x_1, x_2, \ldots, x_n \right) = 0\\
\noalign{\medskip} f_2 \left( x_1, x_2, \ldots, x_n \right) = 0
\end{array}
\right.
\]
defines a surface of dimension $n - 2$, and so on. Since equations are homogeneous, these surfaces posess a \emph{conic} shape: if $\left( x_1, x_2, \ldots, x_n \right)$ is a solution, then $const \cdot \left( x_1, x_2, \ldots, x_n \right)$ is a solution as well. It is convenient to consider solutions, which differ only by rescaling, as equivalent. Such an identification of points in ${\mathbb C}^n$
$$\left( x_1, x_2, \ldots, x_n \right) \sim const \cdot \left( x_1, x_2, \ldots, x_n \right)$$
is called \emph{the projective equivalence} or simply projectivisation. The set of equivalence classes, that is, the set of all one-dimensional subspaces in ${\mathbb C}^n$, is usually denoted as ${\mathbb CP}^{n-1}$. Projectively, a system of $k$ polynomials in general position has $n - k - 1$ dimensional space of solutions.

\smallskip

However, for \emph{special} values of coefficients a system can have more solutions, than it has in the general position. This is a very important phenomenon called \emph{degeneration}. A system of $k$ homogeneous polynomials, which has more than a $(n - k - 1)$-dimensional space of solutions, is called a degenerate system. In this paper, we will be interested only in systems with $k = n$: in this case $n - k - 1 = -1$, meaning that for generic values of coefficients the system does not have projective solutions at all. In other words, a system of $n$ homogeneous polynomials in $n$ variables always has a trivial solution
$$\left( x_1, x_2, \ldots, x_n \right) = \left( 0,0,\ldots,0 \right)$$
but generally it does not have non-trivial solutions. To have a non-trivial solution, coefficients of a system
\[
\left\{
\begin{array}{lll}
f_1 \left( x_1, x_2, \ldots, x_n \right) = 0\\
\noalign{\medskip}f_2 \left( x_1, x_2, \ldots, x_n \right) = 0\\
\noalign{\medskip}\ldots \\
\noalign{\medskip}f_n \left( x_1, x_2, \ldots, x_n \right) = 0\\
\end{array}
\right.
\]
should satisfy one algebraic constraint
$$ R_{r_1, r_2, \ldots, r_n} \big\{ f_1, f_2, \ldots, f_n \big\} = 0 $$
where $R_{r_1, r_2, \ldots, r_n}$ is an irreducible polynomial function called \emph{resultant}, depending on coefficients of the non-linear system under consideration. The subscripts $r_1, r_2, \ldots, r_n$ indicate the degrees of equations. It is often convenient to omit the subscripts, denoting the resultant as $$ R\big\{f_1, f_2, \ldots, f_n\big\} $$ or even simply as $ R(f) $. The existence of such a function, which defines a solvability condition for non-linear systems, can be proved \cite{GKZ}. It can be also proved that resultant is unique up to overall constant factor (which is not important for most applications). Another classical theorem is that resultant is a homogeneous polynomial in coefficients of each equation $f_i$ of degree $$ d_i = r_1 r_2 \ldots r_n \ \dfrac{1}{r_i} $$ therefore total degree in $f$ is $$ d = d_1 + \ldots + d_n = r_1 r_2 \ldots r_n \left( \dfrac{1}{r_1} + \dfrac{1}{r_2} + \ldots + \dfrac{1}{r_n} \right) $$ Unfortunately, most of these proofs are non-constructive, i.e. they prove the property, but give no explicit formula for the resultant.

\smallskip

When all equations are linear, then $(r_1, r_2, \ldots, r_n) = (1,1, \ldots, 1)$ and resultant is just a determinant:
$$R_{\underbrace{1,1,\ldots,1}_{n}} \big( f \big) = det_{n \times n} \big( f \big)$$
Therefore, linear algebra is a natural part of non-linear algebra. This case of linear polynomials is studied in great detail, and an equally detailed description should exist for all degrees. The point is that general conditions of solvability, corresponding to non-linear algebraic equations, are in a sence equally simple as determinants, but far more interesting, especially in view of their potential applications.

\section{Recursive relation between resultants}

An interesting property of resultants is existence of recursive relations between them. Such relations establish connection between resultants of non-linear systems with different number of variables and/or degree of equations. The central role in this paper is played by the following relation between $R_{r_1,r_2,\ldots,r_n}$ and $R_{1,r_2,\ldots,r_n}$:

\paragraph{Proposition I.} Resultant of $n$ homogeneous polynomials of degrees $r_1, r_2, \ldots, r_n$
\be
\left\{
\begin{array}{lll}
(x_1)^{r_1} - f_1 \left( {\vec x} \right) = 0\\
\noalign{\medskip}(x_2)^{r_2} - f_2 \left( {\vec x} \right) = 0\\
\noalign{\medskip}\ldots \\
\noalign{\medskip}(x_n)^{r_n} - f_n \left( {\vec x} \right) = 0\\
\end{array}
\right.
\label{System}
\ee
is expressed through a simpler resultant of $n$ homogeneous polynomials of degrees $1, r_2, \ldots, r_n$
\be
\left\{
\begin{array}{lll}
x_1 - {\vec v} {\vec x} = 0\\
\\
(x_2)^{r_2} - f_2 \left( {\vec x} \right) = 0\\
\noalign{\medskip}\ldots \\
\noalign{\medskip}(x_n)^{r_n} - f_n \left( {\vec x} \right) = 0\\
\end{array}
\right.
\label{LSystem}
\ee
in the following way:
$$
\left. \log R_{r_1 r_2 \ldots r_n} \left\{
\begin{array}{ccc}
(x_1)^{r_1} - f_1 \left( {\vec x} \right) \\
\noalign{\medskip}(x_2)^{r_2} - f_2 \left( {\vec x} \right) \\
\noalign{\medskip}\ldots \\
\noalign{\medskip}(x_n)^{r_n} - f_n \left( {\vec x} \right) \\
\end{array}
\right\} \ = \ \sum\limits_{k = 0}^{\infty} \ \dfrac{r_1}{(r_1 k)!} \ f_1 \left( \dfrac{\partial }{\partial {\vec v}} \right)^k \ \log R_{1 r_2 \ldots r_n} \left\{
\begin{array}{ccc}
x_1 - {\vec v} {\vec x} \\
\\
(x_2)^{r_2} - f_2 \left( {\vec x} \right) \\
\noalign{\medskip}\ldots \\
\noalign{\medskip}(x_n)^{r_n} - f_n \left( {\vec x} \right) \\
\end{array}
\right\} \ \right|_{v = 0}
$$
where ${\vec v} {\vec x}$ is a scalar product $v_1 x_1 + v_2 x_2 + \ldots + v_n x_n$.

\paragraph{ Proof.} Consider a system of $n - 1$ polynomials, which is a common sub-system of (\ref{System}) and (\ref{LSystem}):
\be
\left\{
\begin{array}{lll}
(x_2)^{r_2} - f_2 \left( {\vec x} \right) = 0\\
\noalign{\medskip}\ldots \\
\noalign{\medskip}(x_n)^{r_n} - f_n \left( {\vec x} \right) = 0\\
\end{array}
\right.
\label{SubSystem}
\ee
Suppose that this system is in general position. Then, projectively it has a 0-dimensional space of solutions, that means, a finite number $N$ of one-dimensional subspaces in ${\mathbb C}^n$. Actually, $N = r_2 r_3 \ldots r_n$, but this fact will not be used. Let us select one arbitrary vector on each one-dimensional subspace. \pagebreak We obtain $N$ vectors
$${\vec \Lambda^{(1)}}, {\vec \Lambda^{(2)}}, \ldots, {\vec \Lambda^{(N)}}$$
where ${\vec \Lambda^{(i)}}$ is a vector on the $i$-th one-dimensional subspace, with components $\Lambda^{(i)}_j, \ i = 1 \ldots N, \ j = 1 \ldots n $. These vectors will be called \emph{roots} of (\ref{SubSystem}). Roots are important, because resultant of a system
\be
\left\{
\begin{array}{lll}
\phi \left( {\vec x} \right) = 0 \\
\noalign{\medskip}(x_2)^{r_2} - f_2 \left( {\vec x} \right) = 0\\
\noalign{\medskip}\ldots \\
\noalign{\medskip}(x_n)^{r_n} - f_n \left( {\vec x} \right) = 0\\
\end{array}
\right.
\label{PhiSystem}
\ee
where $\phi$ is an arbitrary polynomial, is simply expressed in terms of roots:
$$R \left\{
\begin{array}{ccc}
\phi \left( {\vec x} \right) \\
\noalign{\medskip}(x_2)^{r_2} - f_2 \left( {\vec x} \right) \\
\noalign{\medskip}\ldots \\
\noalign{\medskip}(x_n)^{r_n} - f_n \left( {\vec x} \right) \\
\end{array}
\right\} = C \cdot \phi \left( {\vec \Lambda^{(1)}} \right) \phi \left( {\vec \Lambda^{(2)}} \right) \ldots \phi \left( {\vec \Lambda^{(N)}}  \right)$$
This statement is known as \emph{the Poisson product formula} \cite{GKZ}. It is valid for the following reason. By definition of solvability, (\ref{PhiSystem}) is solvable if and only if one of values $\phi \left( {\vec \Lambda^{(i)}} \right)$ vanishes. What is the same, resultant of (\ref{PhiSystem}) vanishes if and only if a product of values $\phi\left( {\vec \Lambda^{(i)}} \right)$ vanishes. Since resultant is a polynomial, it must be divisible on the product of values $\phi\left( {\vec \Lambda^{(i)}} \right)$. Therefore, Poisson product formula is valid. The coefficient $C$ depends on the normalisation of roots, and we assume that \medskip roots are normalised in such a way, that $C = 1$.

Let us apply the Poisson product formula to the systems (\ref{System}) and (\ref{LSystem}). If we select $$\phi(x) = (x_1)^{r_1} - f_1({\vec x})$$ then we obtain the resultant of (\ref{System}):
$$ R_{r_1 r_2 \ldots r_n} \left\{
\begin{array}{ccc}
(x_1)^{r_1} - f_1 \left( {\vec x} \right) \\
\noalign{\medskip}(x_2)^{r_2} - f_2 \left( {\vec x} \right) \\
\noalign{\medskip}\ldots \\
\noalign{\medskip}(x_n)^{r_n} - f_n \left( {\vec x} \right) \\
\end{array}
\right\} = \prod\limits_{i = 1}^{N} \phi \left( {\vec \Lambda^{(i)}} \right) = \prod\limits_{i = 1}^{N} \left[ \left(\Lambda^{(i)}_1\right)^{r_1} - f_1\left({\vec \Lambda^{(i)}}\right) \right] $$
Consequently, logaripthm of the resultant of (\ref{System}) is given by
$$
\log \ R_{r_1 r_2 \ldots r_n} \left\{
\begin{array}{ccc}
(x_1)^{r_1} - f_1 \left( {\vec x} \right) \\
\noalign{\medskip}(x_2)^{r_2} - f_2 \left( {\vec x} \right) \\
\noalign{\medskip}\ldots \\
\noalign{\medskip}(x_n)^{r_n} - f_n \left( {\vec x} \right) \\
\end{array}
\right\} = r_1 \sum\limits_{i = 1}^{N} \log \Lambda^{(i)}_1 + \sum\limits_{i = 1}^{N} \log \ \left[ 1 - f_1 \left(\dfrac{{\vec \Lambda^{(i)}}}{ \Lambda^{(i)}_1} \right) \right]
$$
Let us make the series expansion, using $ \ \log \ \left( 1 - x \right) = - \sum\limits_{k = 1}^{\infty} \dfrac{x^k}{k} $. Then we obtain
\be
\log \ R_{r_1 r_2 \ldots r_n} \left\{
\begin{array}{ccc}
(x_1)^{r_1} - f_1 \left( {\vec x} \right) \\
\noalign{\medskip}(x_2)^{r_2} - f_2 \left( {\vec x} \right) \\
\noalign{\medskip}\ldots \\
\noalign{\medskip}(x_n)^{r_n} - f_n \left( {\vec x} \right) \\
\end{array}
\right\} = r_1 \sum\limits_{i = 1}^{N} \log \Lambda^{(i)}_1 + \sum\limits_{i = 1}^{N} \sum\limits_{k = 1}^{\infty} \dfrac{-1}{k} f_1 \left( {\vec \lambda^{(i)}} \right)^k
\label{x1}
\ee
where ${\vec \lambda^{(i)}} = \dfrac{{\vec \Lambda^{(i)}}}{ \Lambda^{(i)}_1}$. In complete analogy, if we select $\phi(x) = x_1 - {\vec x} {\vec v}$ then we obtain
\be
\log \ R_{1 r_2 \ldots r_n} \left\{
\begin{array}{ccc}
x_1 - {\vec v} {\vec x} \\
\\
(x_2)^{r_2} - f_2 \left( {\vec x} \right) \\
\noalign{\medskip}\ldots \\
\noalign{\medskip}(x_n)^{r_n} - f_n \left( {\vec x} \right) \\
\end{array}
\right\} = \sum\limits_{i = 1}^{N} \log \Lambda^{(i)}_1 + \sum\limits_{i = 1}^{N} \sum\limits_{k = 1}^{\infty} \dfrac{-1}{k} \left( {\vec \lambda^{(i)}} {\vec v} \right)^k
\label{x2}
\ee
To establish a relation between (\ref{x1}) and (\ref{x2}), we need to find a differential operator
$$ \sum\limits_{k = 1}^{\infty} p_k f \left( \dfrac{\partial }{\partial {\vec v}} \right)^k $$
which transforms the series
$$\sum\limits_{k = 1}^{\infty} \dfrac{-1}{k} \left( {\vec \lambda} {\vec v} \right)^k$$
into the series
$$\sum\limits_{k = 1}^{\infty} \dfrac{-1}{k} f \left( {\vec \lambda} \right)^k$$
This is easy to do: for any $p_k$ and $c_m$ there is an identity
\be
\left. \left[ \sum\limits_{k = 0}^{\infty} p_k f \left( \dfrac{\partial }{\partial {\vec v}} \right)^k \right] \cdot \sum\limits_{m = 0}^{\infty} c_m \left( {\vec \lambda} {\vec v} \right)^m \right|_{v = 0} = \sum\limits_{k = 0}^{\infty} (k r)! p_{k} c_{k r} \ f \left( {\vec \lambda} \right)^k \label{Hid}
\ee
where $r$ is the degree of $f$. We search for coefficients $p_k$ satisfying $(k r)! p_{k} c_{k r} = -1/k$, having in mind that $c_m = -1/m$; this immediately leads to
$$p_{k} = \dfrac{r}{(k r)!}$$
By substituting (\ref{Hid}) into (\ref{x1}), we obtain
\[
\begin{array}{ccc}
\log \ R \left\{
\begin{array}{ccc}
(x_1)^{r_1} - f_1 \left( {\vec x} \right) \\
\noalign{\medskip}(x_2)^{r_2} - f_2 \left( {\vec x} \right) \\
\noalign{\medskip}\ldots \\
\noalign{\medskip}(x_n)^{r_n} - f_n \left( {\vec x} \right) \\
\end{array}
\right\} = \left. r_1 \sum\limits_{i = 1}^{N} \log \Lambda^{(i)}_1 + \left[ \sum\limits_{k = 1}^{\infty} \dfrac{r_1}{(k r_1)!} f_1 \left( \dfrac{\partial }{\partial {\vec v}} \right)^k \right] \cdot \log R \left\{
\begin{array}{ccc}
x_1 - {\vec v} {\vec x} \\
\\
(x_2)^{r_2} - f_2 \left( {\vec x} \right) \\
\noalign{\medskip}\ldots \\
\noalign{\medskip}(x_n)^{r_n} - f_n \left( {\vec x} \right) \\
\end{array}
\right\} \ \right|_{v = 0}
\end{array}
\]
or, what is the same,
\[
\begin{array}{ccc}
\log \ R_{r_1 r_2 \ldots r_n} \left\{
\begin{array}{ccc}
(x_1)^{r_1} - f_1 \left( {\vec x} \right) \\
\noalign{\medskip}(x_2)^{r_2} - f_2 \left( {\vec x} \right) \\
\noalign{\medskip}\ldots \\
\noalign{\medskip}(x_n)^{r_n} - f_n \left( {\vec x} \right) \\
\end{array}
\right\} = \left. \left[ \sum\limits_{k = 0}^{\infty} \dfrac{r_1}{(k r_1)!} f_1 \left( \dfrac{\partial }{\partial {\vec v}} \right)^k \right] \cdot \log R_{1 r_2 \ldots r_n} \left\{
\begin{array}{ccc}
x_1 - {\vec v} {\vec x} \\
\\
(x_2)^{r_2} - f_2 \left( {\vec x} \right) \\
\noalign{\medskip}\ldots \\
\noalign{\medskip}(x_n)^{r_n} - f_n \left( {\vec x} \right) \\
\end{array}
\right\} \ \right|_{v = 0}
\end{array}
\]
The proposition is proved. As we see, Poisson product formula can be used to derive explicit relations between resultants, written not in terms of roots, but in terms of coefficients of non-linear equations under consideration. We will see below, that these relations provide a lot of information about resultants: in fact, enough information to calculate them completely.

\section{Generalized identity Log Det = Trace Log}

\paragraph{Proposition II.} Logarithm of the resultant of $n$ homogeneous polynomials of degrees $r_1,r_2,\ldots,r_n$
\[
\left\{
\begin{array}{lll}
(x_1)^{r_1} - f_1 \left( {\vec x} \right) = 0\\
\noalign{\medskip}(x_2)^{r_2} - f_2 \left( {\vec x} \right) = 0\\
\noalign{\medskip}\ldots \\
\noalign{\medskip}(x_n)^{r_n} - f_n \left( {\vec x} \right) = 0\\
\end{array}
\right.
\]
is given by the explicit formula
\be
\log \ R_{r_1 r_2 \ldots r_n} \left\{
\begin{array}{ccc}
(x_1)^{r_1} - f_1 \left( {\vec x} \right) \\
\noalign{\medskip}(x_2)^{r_2} - f_2 \left( {\vec x} \right) \\
\noalign{\medskip}\ldots \\
\noalign{\medskip}(x_n)^{r_n} - f_n \left( {\vec x} \right) \\
\end{array}
\right\} = \left. S_{r_1} \! \left({\hat f}_1 \right) S_{r_2} \! \left({\hat f}_2 \right) \ldots S_{r_n} \! \left({\hat f}_n \right) \cdot \log \ \det(I - A) \right|_{A = 0}
\label{LogRezEx2}
\ee
where $A$ is an auxillary $n \times n$ matrix, $S_r(z)$ denotes the series $$S_r(z) = \sum\limits_{k = 0}^{\infty} \dfrac{r}{(r k)!} z^k$$ and ${\hat f}_1, {\hat f}_2, \ldots, {\hat f}_n$ are differential operators
\[
\begin{array}{ccc}
{\hat f_1} =  f_1 \left( \dfrac{\partial}{\partial A_{11}}, \dfrac{\partial}{\partial A_{12}}, \ldots, \dfrac{\partial}{\partial A_{1n}} \right) \\
\noalign{\medskip} {\hat f_2} =  f_2 \left( \dfrac{\partial}{\partial A_{21}}, \dfrac{\partial}{\partial A_{22}}, \ldots, \dfrac{\partial}{\partial A_{2n}} \right) \\
\noalign{\medskip} \ldots \\
\noalign{\medskip} {\hat f_n} =  f_n \left( \dfrac{\partial}{\partial A_{n1}}, \dfrac{\partial}{\partial A_{n2}}, \ldots, \dfrac{\partial}{\partial A_{nn}} \right) \\
\end{array}
\]
associated with polynomials $f_1, f_2, \ldots, f_n$. More explicitly,

$$
\log \ R_{r_1 r_2 \ldots r_n} \left\{
\begin{array}{ccc}
(x_1)^{r_1} - f_1 \left( {\vec x} \right) \\
\noalign{\medskip}(x_2)^{r_2} - f_2 \left( {\vec x} \right) \\
\noalign{\medskip}\ldots \\
\noalign{\medskip}(x_n)^{r_n} - f_n \left( {\vec x} \right) \\
\end{array}
\right\} = \left. \left[ \ \sum\limits_{k = 0}^{\infty} \dfrac{r_1}{(r_1 k)!} ({\hat f_1})^{k} \ \right] \ldots \left[ \ \sum\limits_{k = 0}^{\infty} \dfrac{r_n}{(r_n k)!} ({\hat f_n})^{k} \ \right] \cdot \log \ \det(I - A) \ \right|_{A = 0} $$

\paragraph{Proof.} This statement is easily proved by subsequent application of Proposition I, which allows to decrease degree of each of the equations. To begin with, we decrease the degree of the first equation
$$
\left. \log R \left\{
\begin{array}{ccc}
(x_1)^{r_1} - f_1 \left( {\vec x} \right) \\
\noalign{\medskip}(x_2)^{r_2} - f_2 \left( {\vec x} \right) \\
\noalign{\medskip}\ldots \\
\noalign{\medskip}(x_n)^{r_n} - f_n \left( {\vec x} \right) \\
\end{array}
\right\} \ = \ S_{r_1} \left[ f_1 \left( \dfrac{\partial }{\partial {\vec v}} \right) \right] \ \log R \left\{
\begin{array}{ccc}
x_1 - {\vec v} {\vec x} \\
\\
(x_2)^{r_2} - f_2 \left( {\vec x} \right) \\
\noalign{\medskip}\ldots \\
\noalign{\medskip}(x_n)^{r_n} - f_n \left( {\vec x} \right) \\
\end{array}
\right\} \ \right|_{v = 0}
$$
then, we decrease the degree of the second equation
$$
\left. \log R \left\{
\begin{array}{ccc}
(x_1)^{r_1} - f_1 \left( {\vec x} \right) \\
\noalign{\medskip}(x_2)^{r_2} - f_2 \left( {\vec x} \right) \\
\noalign{\medskip}\ldots \\
\noalign{\medskip}(x_n)^{r_n} - f_n \left( {\vec x} \right) \\
\end{array}
\right\} \ = \ S_{r_1} \left[ f_1 \left( \dfrac{\partial }{\partial {\vec v}} \right) \right]
S_{r_2} \left[ f_2 \left( \dfrac{\partial }{\partial {\vec w}} \right) \right]
\ \log R \left\{
\begin{array}{ccc}
x_1 - {\vec v} {\vec x} \\
\\
x_2 - {\vec w} {\vec x} \\
\noalign{\medskip}\ldots \\
\noalign{\medskip}(x_n)^{r_n} - f_n \left( {\vec x} \right) \\
\end{array}
\right\} \ \right|_{v,w = 0}
$$
and so on. After repeating this procedure $n$ times, all equations become linear:
$$
\left. \log R \left\{
\begin{array}{ccc}
(x_1)^{r_1} - f_1 \left( {\vec x} \right) \\
\noalign{\medskip}(x_2)^{r_2} - f_2 \left( {\vec x} \right) \\
\noalign{\medskip}\ldots \\
\noalign{\medskip}(x_n)^{r_n} - f_n \left( {\vec x} \right) \\
\end{array}
\right\} \ = \ \prod\limits_{i = 1}^{n} S_{r_i} \left[ f_i \left( \dfrac{\partial }{\partial {\vec v^{(i)}}} \right) \right] \
\log R \left\{
\begin{array}{ccc}
x_1 - {\vec v^{(1)}} {\vec x} \\
\\
x_2 - {\vec v^{(2)}} {\vec x} \\
\noalign{\medskip}\ldots \\
\noalign{\medskip}x_2 - {\vec v^{(n)}} {\vec x} \\
\end{array}
\right\} \ \right|_{{\vec v^{(1)}},\ldots,{\vec v^{(n)}} = 0}
$$Here, ${\vec v^{(1)}},\ldots,{\vec v^{(n)}}$ are $n$ auxillary vectors. As we see, the resultant of $n$ polynomials of degrees $r_1,r_2,\ldots,r_n$ is expressed through the determinant of $n$ linear polynomials. If we denote
$A_{ij} = v^{(i)}_j$ and
$${\hat f}_i = f_i \left( \dfrac{\partial }{\partial {\vec v^{(i)}}} \right) = f_i \left( \dfrac{\partial}{\partial A_{i1}}, \dfrac{\partial}{\partial A_{i2}}, \ldots, \dfrac{\partial}{\partial A_{in}} \right)$$
then we finally obtain
$$
\log \ R_{r_1 r_2 \ldots r_n} \left\{
\begin{array}{ccc}
(x_1)^{r_1} - f_1 \left( {\vec x} \right) \\
\noalign{\medskip}(x_2)^{r_2} - f_2 \left( {\vec x} \right) \\
\noalign{\medskip}\ldots \\
\noalign{\medskip}(x_n)^{r_n} - f_n \left( {\vec x} \right) \\
\end{array}
\right\} = \left. \left[ \ \sum\limits_{k = 0}^{\infty} \dfrac{r_1}{(r_1 k)!} ({\hat f_1})^{k} \ \right] \ldots \left[ \ \sum\limits_{k = 0}^{\infty} \dfrac{r_n}{(r_n k)!} ({\hat f_n})^{k} \ \right] \cdot \log \ \det(I - A) \ \right|_{A = 0} $$
and so the proposition is proved. Since it explicitly describes the series expansion of $\log \ R_{r_1, \ldots, r_n}(I - f)$, it is natural to call it the generalized Log Det = Trace Log identity.

\section{Generalized shift operator}

Differential operators, which appear in eq. (\ref{LogRezEx2}), are expressed in terms of the following special function:
$$ {\cal S}_{r}(z) = \sum\limits_{k = 0}^{\infty} \dfrac{r}{(k r)!} z^k = r \left( 1 + \dfrac{z}{r!} + \dfrac{z^2}{(2 r)!} + \dfrac{z^3}{(3 r)!} + \ldots \right) $$
so that
\[
\begin{array}{cc}
{\cal S}_{1}(z) = 1 + z + \dfrac{1}{2} z^2 + \dfrac{1}{6} z^3 + \ldots \\
\\
{\cal S}_{2}(z) = 2 + z + \dfrac{1}{12} z^2 + \dfrac{1}{360} z^3 + \ldots \\
\\
{\cal S}_{3}(z) = 3 + \dfrac{1}{2} z + \dfrac{1}{240} z^2 + \dfrac{1}{120960} z^3 + \ldots \\
\end{array}
\]
and so on. Operators ${\cal S}_{r_i}\left({\hat f}_i\right)$ establish an explicit relation (\ref{LogRezEx2}) between resultants and determinants. For this reason, their properties deserve to be further investigated. It is very interesting, that such operators naturally appear in the study of non-linear systems.
\pagebreak
Note, that for $r = 1$ we have ${\cal S}_{1}(z) = \exp \left( z \right)$. Therefore, if all equations are linear, corresponding operators are nothing but shift operators
$$ {\cal S}_{1}\left({\hat f}_i\right) = \exp \left( {\hat f}_i \right) = \exp \left( f_{i1} \dfrac{\partial}{\partial A_{i1}} + f_{i2} \dfrac{\partial}{\partial A_{i2}} + \ldots + f_{in} \dfrac{\partial}{\partial A_{in}} \right) $$
According to the elementary shift identity
$$\exp \left( f_{ij} \dfrac{\partial}{\partial A_{ij}} \right) \cdot \phi( A ) = \phi( A + f )$$
where $\phi$ is \emph{any} function of matrix $A$, the role of these operators is to substitute $f_{ij}$ in place of every $A_{ij}$:
$$\left. {\cal S}_{1}\left({\hat f_1}\right) {\cal S}_{1}\left({\hat f_2}\right) \ldots {\cal S}_{1}\left({\hat f_n}\right) \cdot \log \ \det(I - A) \right|_{A = 0} = \left. \det(I - A - f) \right|_{A = 0} = \det(I - f)$$

$$\left. {\cal S}_{1}\left({\hat f_1}\right) {\cal S}_{1}\left({\hat f_2}\right) \ldots {\cal S}_{1}\left({\hat f_n}\right) \cdot {\rm tr} \ A^{k} \right|_{A = 0} = \left. {\rm tr} \ (A + f)^{k} \right|_{A = 0} = {\rm tr} \ f^{k}$$
However, for non-linear polynomials action of these operators is no longer that simple. Instead, they provide a non-trivial generalization of the shift operator, which maps determinants into resultants

$$
\left. S_{r_1} \! \left({\hat f}_1 \right) S_{r_2} \! \left({\hat f}_2 \right) \ldots S_{r_n} \! \left({\hat f}_n \right) \cdot \log \ \det(I - A) \right|_{A = 0} = \log \ R_{r_1 r_2 \ldots r_n} \left\{
\begin{array}{ccc}
(x_1)^{r_1} - f_1 \left( {\vec x} \right) \\
\noalign{\medskip}(x_2)^{r_2} - f_2 \left( {\vec x} \right) \\
\noalign{\medskip}\ldots \\
\noalign{\medskip}(x_n)^{r_n} - f_n \left( {\vec x} \right) \\
\end{array}
\right\}
$$
and traces of powers into their non-linear analogues:

$$
\left. S_{r_1} \! \left({\hat f}_1 \right) S_{r_2} \! \left({\hat f}_2 \right) \ldots S_{r_n} \! \left({\hat f}_n \right) \cdot {\rm tr} A^{k} \ \right|_{A = 0} = T_{k}(f)
$$
Function ${\cal S}_r(z)$ has many different representations, which can be useful under different circumstances. We list just a few representations. First of all, ${\cal S}_r(z)$ belongs to the class of hypergeometric functions, defined by the following series
$$
{}_rF_s(a_1 , \dotsc , a_r; b_1 , \dotsc , b_s; z) \equiv \sum_{k=0}^\infty
\frac{(a_1)_k \dotsb (a_r)_k}{(b_1)_k \dotsb (b_s)_k} \frac{z^k}{k!}
$$
where
$$
(a)_k \equiv \frac{\Gamma(a+k)}{\Gamma(a)} = a(a+1) \dotsb (a+k-1).
$$
Our function is equal to
$${\cal S}_r(z) = {}_0F_{r-1}\left(\dfrac{1}{r},\dfrac{2}{r},\ldots,\dfrac{r-1}{r}; \dfrac{z}{r^r}\right)$$
because
$$(r k)! = r^{r k} \ k! \ \left(\dfrac{1}{r}\right)_{k} \left(\dfrac{2}{r}\right)_{k} \ldots \left(\dfrac{r-1}{r}\right)_{k}$$
In particular,
\[
\begin{array}{ccc}
{\cal S}_{1}(z)   = {}_0F_0\left(z\right) = \exp \left( z \right) \\
   \\
{\cal S}_{2}(z)   = {}_0F_1 \left(\dfrac{1}{2}; \dfrac{z}{4}\right) \\
  \\
{\cal S}_{3}(z)   = {}_0F_2 \left(\dfrac{1}{3}, \dfrac{2}{3}; \dfrac{z}{27}\right) \\
  \\
{\cal S}_{4}(z)   = {}_0F_3 \left(\dfrac{1}{4}, \dfrac{2}{4}, \dfrac{3}{4}; \dfrac{z}{256}\right)
\end{array}
\]
It deserves noting, that ${\cal S}_r(z)$ is expressable through elementary functions not only for $r = 1$, but for any $r$:
\[
\begin{array}{ccc}
{\cal S}_{1}(z) = \exp \left( z \right) \\
\\
{\cal S}_{2}(z) = \exp \left( \sqrt{z} \right) + \exp \left( - \sqrt{z} \right)\\
\\
{\cal S}_{3}(z) = \exp \left( \sqrt[3]{z} \right) + \exp \left( \omega_3 \ \sqrt[3]{z} \right) + \exp \left( (\omega_3)^2 \ \sqrt[3]{z}  \right), \ \ \ \omega_3 = e^{2 \pi i/3} \\
\\
{\cal S}_{4}(z) = \exp \left( \sqrt[4]{z} \right) + \exp \left( \omega_4 \ \sqrt[4]{z} \right) + \exp \left( (\omega_4)^2 \ \sqrt[4]{z}  \right) + \exp \left( (\omega_4)^3 \ \sqrt[4]{z}  \right), \ \ \ \omega_4 = e^{2 \pi i/4} \\
\end{array}
\]
and so on:
$${\cal S}_{r}(z) = \sum\limits_{p = 0}^{r-1} \exp \left( (\omega_r)^p \sqrt[r]{z} \right), \ \ \ \omega_r = e^{2 \pi i/r} = \cos ( 2 \pi/r) + i \sin ( 2 \pi/r) $$
where $ \omega_r $ is the complex root of unity of order $r$. However, at the present stage of development this expression through elementary functions does not seem to be very helpful, because its particular constituents contain fractional powers of $z$ and associated sign ambiguities. More useful can be various integral representations for ${\cal S}_r(z)$, explicitly involving only integer powers of $z$. For example, using
$$ {}_0F_1 \left(a; z\right) = \dfrac{1}{\Gamma (1 - a)} \int\limits_{0}^{\infty} \dfrac{dt}{t^a} \exp \left( - t + \dfrac{z}{t} \right) $$
$$ {}_0F_2 \left(a,b; z\right) = \dfrac{1}{\Gamma (1 - a)\Gamma (1 - b)} \int\limits_{0}^{\infty} \int\limits_{0}^{\infty} \dfrac{dt_1 dt_2}{t_1^a t_2^{b} } \exp \left( - t_1 - t_2 + \dfrac{z}{t_1 t_2} \right) $$
we express the generalized shift operators as integrals
$$ {\cal S}_{2}({\hat f}) = \dfrac{1}{\Gamma (1/2)} \int\limits_{0}^{\infty} \dfrac{dt}{t^{1/2}} \exp \left( - t + \dfrac{{\hat f}}{t} \right) $$
$$ {\cal S}_{3}({\hat f}) = \dfrac{1}{\Gamma (1/3)\Gamma (2/3)} \int\limits_{0}^{\infty} \int\limits_{0}^{\infty} \dfrac{dt_1 dt_2}{t_1^{1/3} t_2^{2/3} } \exp \left( - t_1 - t_2 + \dfrac{{\hat f}}{t_1 t_2} \right) $$
Note also, that exponents of differential operators are naturally represented as Feynman functional integrals (see \cite{MV} for such free-field representation of the hypergeometric series). In this way, the generalized shift operators can be expressed as functional integrals. Also of interest are much simpler matrix-integral representations, which we present in the following section.

\section{Matrix integral representation}

Relation (\ref{LogRezEx2}) is written in terms of differential operators like
$$\dfrac{ \partial }{\partial A_{ij} }$$
which act on $\log \det(I - A)$. Appearance of such differential operators strongly suggests to rewrite (\ref{LogRezEx2}) in the integral form, using a Fourier transform from variables $A_{ij}$ to conjugate variables (momenta) $P_{ij}$:
\be
 \log \ \det(I - A) = \int dP \ \phi(P) \ e^{ {\rm tr} \left( A P \right) }
\label{HHH}
\ee
The integral here is taken over all entries $P_{ij}$ of a $n \times n$ matrix $P$ with the measure $dP = \prod\limits_{i,j = 1}^{n} dP_{ij}$.
Function $\phi(P)$ is given by the inverse Fourier transform
$$ \phi(P) = \int dQ \ e^{ {\rm tr} \left( P Q \right) } \ \log \det(I - Q) $$
The main advantage of using Fourier transformations is a greatly simplified action of differential operators. In our case, the action of differential operators has the form
$$\dfrac{ \partial }{\partial A_{\alpha \beta} } \log \ \det(I - A) = \int dP \ \phi(P) \ e^{ {\rm tr} \left( A P \right) } \ P_{\alpha \beta} $$
Consequently, (\ref{LogRezEx2}) turns into a matrix integral
$$ \log R_{r_1 r_2 \ldots r_n} \left\{
\begin{array}{ccc}
(x_1)^{r_1} - f_1 \left( {\vec x} \right) \\
\noalign{\medskip}(x_2)^{r_2} - f_2 \left( {\vec x} \right) \\
\noalign{\medskip}\ldots \\
\noalign{\medskip}(x_n)^{r_n} - f_n \left( {\vec x} \right) \\
\end{array}
\right\} \ = \int dP \ \phi(P) \left[ \prod\limits_{i = 1}^{n} \sum\limits_{k = 0}^{\infty} \dfrac{r_i}{(r_i k)!} f_{i}(P)^{k} \right]  $$
where $f_i(P) = f_i(P_{i1},P_{i2},\ldots,P_{in})$. Using the expression for $\phi(P)$, we obtain
$$ \log R_{r_1 r_2 \ldots r_n} \left\{
\begin{array}{ccc}
(x_1)^{r_1} - f_1 \left( {\vec x} \right) \\
\noalign{\medskip}(x_2)^{r_2} - f_2 \left( {\vec x} \right) \\
\noalign{\medskip}\ldots \\
\noalign{\medskip}(x_n)^{r_n} - f_n \left( {\vec x} \right) \\
\end{array}
\right\} \ = \int \int dP dQ \ e^{ {\rm tr} \left( P Q \right) } \ \log \det(I - Q) \ \left[ \prod\limits_{i = 1}^{n} \sum\limits_{k = 0}^{\infty} \dfrac{r_i}{(r_i k)!} f_{i}(P)^{k} \right] $$
This is a matrix integral representation of the logarithm of resultant. The next step in evaluation of this matrix integral is separation of eigenvalues and angular variables. Functions $f_{i}(P)$ depend non-trivially on angular variables, and the corresponding integral can probably be expressed through $SL(n)$ characters by general methods of \cite{MMS}.

Relation (\ref{LogRezEx2}) can be used to derive other integral representations and it would be especially interesting to obtain an integral representation for the resultant itself, not just for its logarithm. This kind of formulas provide new tools for exact (non-perturbative) evaluation of non-Gaussian integrals, which certainly deserves all possible attention.

\section{Resultant is a Schur polynomial}

As a direct application, (\ref{LogRezEx}) produces explicit polynomial formulas for resultants in any dimension.
In order to obtain such formulas, one should pass from logarithms to resultants themselves:
\be
R_{r_1,\ldots,r_n}(I - f) = \exp \left( \left. \left[ \ \sum\limits_{k = 0}^{\infty} \dfrac{r_1}{(r_1 k)!} ({\hat f_1})^{k} \ \right] \ldots \left[ \ \sum\limits_{k = 0}^{\infty} \dfrac{r_n}{(r_n k)!} ({\hat f_n})^{k} \ \right] \cdot \log \ \det(I - A) \ \right|_{A = 0} \right)
\label{Expo}
\ee
Explicit polynomial formulas follow from the expansion of the right hand side into power series in $f_1, f_2, \ldots, f_n$, which is a typical example of Schur expansions. Let us remind, that a \emph{Schur expansion} is the power series expansion of $exp(S(x))$, where $S(x)$ is itself a power series (perhaps, of many variables). When one makes such expansions, one needs to collect terms of equal degree. The simplest example is provided by the \emph{ordinary Schur polynomials} $P_k\left\{ t_i \right\}$, which are defined by
\be
\exp \Big( \sum\limits_{k = 0}^{\infty} t_{k} \phi^k \Big) = \sum\limits_{k = 0}^{\infty} P_{k}\left\{ t_i \right\} \ \phi^k
\label{GenSchur}
\ee
where $t_{0} = 0$. Using $\exp \left( x \right) = \sum\limits_{m = 0}^{\infty} \dfrac{x^m}{m!}$, one can find several Schur polynomials
\[
\begin{array}{cc}
P_0\left\{ t_i \right\} = 1\\
\\
P_1\left\{ t_i \right\} = t_1 \\
\\
P_2\left\{ t_i \right\} = t_{2} + t_{1}^2/2 \\
\\
P_3\left\{ t_i \right\} = t_{3} + t_{1} t_{2} + t_{1}^3/6 \\
\end{array}
\]
as well as the explicit expression for $P_k\left\{ t_i \right\}$:
\be
P_{k}\left\{ t_i \right\} = \sum\limits_{m = 1}^{k} \ \ \sum\limits_{v_1 + v_2 + \ldots + v_m = k} \dfrac{t_{v_1} t_{v_2} \ldots t_{v_m}}{m!}
\label{explSchur}
\ee
where the last sum is taken over all ordered partitions of $k$ into $m$ positive parts, denoted as $v_1, \ldots, v_m$. In complete analogy, if one considers many variables $\phi_{1}, \ldots, \phi_{n}$ instead of one variable $\phi$, then \emph{multi-Schur polynomials} ${\cal P}_{k_1 k_2 \ldots k_n}$ are defined by
\be
\exp \Big( \sum\limits_{k_1, \ldots, k_n = 0}^{\infty} t_{k_1 k_2 \ldots k_n} \phi_{1}^{k_1} \phi_{2}^{k_2} \ldots \phi_{n}^{k_n} \Big) = \sum\limits_{k_1, \ldots, k_n = 0}^{\infty} {\cal P}_{k_1 k_2 \ldots k_n}\left\{ t_{i_1,\ldots,i_n} \right\} \ \phi_{1}^{k_1} \phi_{2}^{k_2} \ldots \phi_{n}^{k_n}
\label{GenMultiSchur}
\ee
where $t_{00\ldots0} = 0$. Several multi-Schur polynomials are
\[
\begin{array}{cc}
{\cal P}_{1, 0}\left\{ t_{ij} \right\} = t_{1, 0}\\
\\
{\cal P}_{2, 0}\left\{ t_{ij} \right\} = t_{2, 0} + t_{1, 0}^2/2\\
\\
{\cal P}_{3, 0}\left\{ t_{ij} \right\} = t_{3, 0} + t_{2, 0} t_{1, 0} + t_{1, 0}^3/6\\
\\
{\cal P}_{2, 1}\left\{ t_{ij} \right\} = t_{2,1}+t_{2,0} t_{0,1}+t_{1,0} t_{1,1}+ t_{1,0}^2 t_{0,1}/2\\
\\
{\cal P}_{3, 1}\left\{ t_{ij} \right\} = t_{3,1}+t_{1,0} t_{2,1}+t_{2,0} t_{1,1}+t_{3,0} t_{0,1}+t_{2,0} t_{0,1} t_{1,0}+ t_{1,0}^2 t_{1,1}/2+ t_{1,0}^3 t_{0,1}/6\\
\\
{\cal P}_{1, 1, 1}\left\{ t_{ijk} \right\} = t_{1,1,1}+t_{1,0,0} t_{0,1,1}+t_{0,1,0} t_{1,0,1}+t_{1,1,0} t_{0,0,1}+t_{0,1,0} t_{0,0,1} t_{1,0,0}\\
\\
{\cal P}_{2, 1, 0}\left\{ t_{ijk} \right\} = t_{2,1,0}+t_{2,0,0} t_{0,1,0}+t_{1,0,0} t_{1,1,0}+t_{1,0,0}^2 t_{0,1,0}/2\\
\end{array}
\]
and so on, and the explicit expression for ${\cal P}_{k_1 k_2 \ldots k_n}$ is
\be
P_{{\vec k}}\left\{ t_{\vec i} \right\} = \sum\limits_{m = 1}^{k_1 + k_2 + \ldots + k_n} \ \ \sum\limits_{{\vec v_1} + {\vec v_2} + \ldots + {\vec v_m} = {\vec k} } \dfrac{t_{{\vec v_1}} t_{{\vec v_2}} \ldots t_{{\vec v_m}}}{m!}
\label{explMultiSchur}
\ee
As one can see, it is very similar to the explicit expression for ordinary Schur polynomials. The sum here is taken over all ordered partitions of a a vector ${\vec k} = (k_1, k_2, \ldots, k_n)$ into $m$ vectors, denoted as ${\vec v_1}, \ldots, {\vec v_m}$. Such an ordered partition is a way of writing a $n$-vector with integer components as a sum of $n$-vectors with integer components, where the order of the items is significant.

There are two ways to produce explicit polynomial formulas for multidimensional resultants. From the first point of view, the terms in (\ref{Expo}) are graded by degrees in $f_1, f_2, \ldots, f_n$
\be
R_{r_1,\ldots,r_n}(I - f) = \exp \left( \sum\limits_{k_1, \ldots, k_n = 0}^{\infty} - T_{k_1, k_2, \ldots, k_n} (f) \right)
\label{PP1}
\ee
where the graded components
$$
T_{k_1, k_2, \ldots, k_n} (f) = \left. r_1 r_2 \ldots r_n \ \dfrac{ ({\hat f_1})^{k_1}}{(r_1 k_1)!} \ldots \dfrac{ ({\hat f_n})^{k_n}}{(r_n k_n)!} \ \cdot \dfrac{ tr A^{r_1 k_1 + \ldots + r_n k_n} } {r_1 k_1 + \ldots + r_n k_n } \ \right|_{A = 0}
$$
are homogeneous polynomials in each $f_i$ of degree $k_i$. Since $R_{r_1,r_2,\ldots,r_n}\big\{ f_1, f_2, \ldots, f_n \big\}$ is homogeneous in each of $f_i$ of degree $d_i$ (as mentioned in s. 2), it can be obtained by collecting terms of that degrees in the expansion of the right hand side of (\ref{PP1}) -- which means that resultant is given by a multi-Schur polynomial
\[
\boxed{
\begin{array}{ccc}
\\ \nonumber
R_{r_1,r_2,\ldots,r_n}(f) = (-1)^{d_1 + d_2 + \ldots + d_n} {\cal P}_{d_1,d_2,\ldots,d_n} \Big\{ - T_{k_1, k_2, \ldots, k_n} (f) \Big\} \\ \nonumber
\end{array}
}
\]
Another way to do is to consider a rougher grading -- by total degree in $f$, neglecting the difference between $f_1, f_2, \ldots, f_n$. From this point of view, terms in the right hand side are graded as
$$ R_{r_1,\ldots,r_n}(I - f) = \exp \left( \sum_{k = 0}^{\infty} \dfrac{-1}{k} T_{k}(f) \right) $$
where the factor of $\dfrac{1}{k}$ is by convention (for similarity with linear algebra). The graded components
\be
\dfrac{T_{k}(f)}{k} = \sum\limits_{k_1 + \ldots + k_n = k} \ T_{k_1, k_2, \ldots, k_n} (f) = \sum\limits_{k_1 + k_2 + \ldots + k_n = k} \left. \ \dfrac{r_1 ({\hat f_1})^{k_1}}{(r_1 k_1)!} \ldots \dfrac{r_n ({\hat f_n})^{k_n}}{(r_n k_n)!} \ \cdot \dfrac{ tr A^{r_1 k_1 + \ldots + r_n k_n} } {r_1 k_1 + \ldots + r_n k_n } \ \right|_{A = 0}
\label{PP2}
\ee
are homogeneous polynomials in $f$ of degree $k$ (but are no longer homogeneous in each $f_i$).

\smallskip

It is these quantities $T_{k}(f)$, which we have mentioned in the introduction as \emph{traces} for non-linear systems. Respectively, multigraded quantities $T_{k_1, k_2, \ldots, k_n}$ can be considered as more elementary building blocks for these traces. Since $R_{r_1,\ldots,r_n}(f)$ is a homogeneous polynomial in $f$ of degree $d$ (as mentioned in s. 2), it can be obtained by collecting terms of that degree in the expansion of the right hand side of (\ref{PP2}) -- which means that resultant is given by a Schur polynomial
\[
\boxed{
\begin{array}{ccc}
\\ \nonumber
R_{r_1,\ldots,r_n}(f) = (-1)^d P_{d} \left\{ \dfrac{- T_{k}(f)}{k} \right\} \\ \nonumber
\end{array}
}
\]
Both formulas are explicit, polynomial and have a straightforward computer implementation. The first formula works somewhat faster, than the second. They allow to calculate a resultant for any $n$ and any $r_1,r_2,\ldots,r_n$. In low dimensions and for low degrees, more simple formulas can be written in the framework of the Cayley approach; however, in higher dimensions our formulas are expected to be more efficient.

Another important issue, concerning Schur polynomials, is that these polynomials satisfy an infinite number of differential equations
$$\dfrac{\partial }{\partial t_{i} } P_{k} = P_{k - i}$$
and their multi- generalizations
$$\dfrac{\partial }{\partial t_{\vec{i}} } {\cal P}_{{\vec k}} = {\cal P}_{{\vec k} - {\vec i}}$$
These relations are known to play an important role in the theory of integrability in matrix models. Since we know, that $R(f)$ is always a Schur polynomial, it should also satisfy certain equations of that type:
$$ \dfrac{ \delta R(f) }{\delta f} \sim \sum\limits_{s = 1}^{d} \dfrac{ \delta P_{d}\left\{ - \dfrac{T_k(f)}{k} \right\} }{\delta T_{s}(f)} \dfrac{ \delta T_{s}(f) }{\delta f} = - \sum\limits_{s = 1}^{d} s P_{d - s}\left\{ - \dfrac{T_k(f)}{k} \right\} \dfrac{ \delta T_{s}(f) }{\delta f} $$
where $T_{k}(f)$ are the traces of a non-linear map $f$. We do not go into details here, and the (possible) relation of resultants to integrability theory remains to be checked.

\section{Resultant is a sum over paths }

Schur polynomials (\ref{explSchur}) and (\ref{explMultiSchur}) have a clear geometrical meaning. An ordinary Schur polynomial
$$
P_{k}\left\{ t_i \right\} = \sum\limits_{m = 1}^{k} \ \ \sum\limits_{v_1 + v_2 + \ldots + v_m = k} \dfrac{t_{v_1} t_{v_2} \ldots t_{v_m}}{m!}
$$
in fact, is a sum over paths $s(i)$ in discrete time $i = 0,1,2,\ldots$ on a one-dimensional discrete space (paths on a lattice), which connect points $0$ and $k$. Each path is uniquely characterised by an ordered partition $v_1, v_2, \ldots, v_m$, where $v_i = s(i) - s(i-1)$ is the velocity at the moment $i$. Therefore,
\[
\begin{array}{lll}
s(0) = 0 \\
\noalign{\medskip} s(1) = v_1 \\
\noalign{\medskip} s(2) = v_1 + v_2 \\
\noalign{\medskip} \ldots \\
\noalign{\medskip} s(m) = v_1 + v_2 + \ldots + v_{m} = k
\end{array}
\]
and the Schur polynomial is given by the sum over all such paths:
$$
P_{k}\left\{ t_i \right\} = \sum\limits_{m = 1}^{k} \ \ \mathop{\sum\limits_{s(i)}}_{s(0) = 0, \ s(m) = k} \dfrac{ t_{v_1} t_{v_2} \ldots t_{v_m} }{m!}
$$
The length $m$ of a path is not smaller than 1, but not greater than $k$. The correspondence between partitions and one-dimensional lattice paths is illustrated by the following picture:
\begin{center}
\includegraphics[width=300pt]{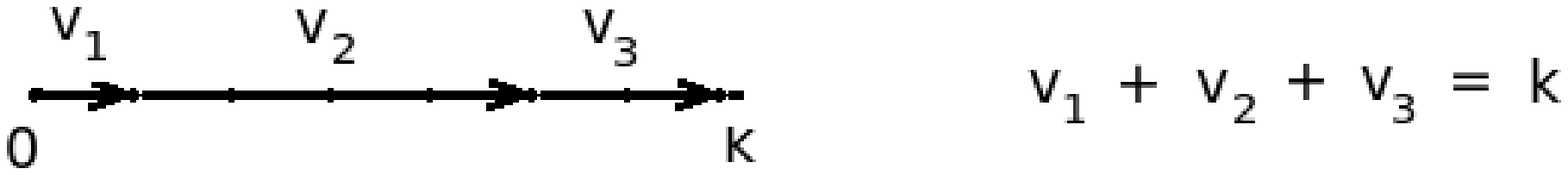}
\end{center}
In complete analogy, a multi-Schur polynomial
$$
P_{{\vec k}}\left\{ t_{\vec i} \right\} = \sum\limits_{m = 1}^{ k_1 + k_2 + \ldots + k_n} \ \ \sum\limits_{{\vec v_1} + {\vec v_2} + \ldots + {\vec v_m} = {\vec k} } \dfrac{t_{{\vec v_1}} t_{{\vec v_2}} \ldots t_{{\vec v_m}}}{m!}
$$
is given by a sum over paths ${\vec s}(i)$ in discrete time $i = 0,1,2,\ldots$ on a $n$-dimensional lattice, beginning in ${\vec 0}$ and ending in ${\vec k}$. Each path is uniquely characterised by an ordered partition ${\vec v}_1, {\vec v}_2, \ldots, {\vec v}_m$, where $${\vec v}_i = {\vec s}(i) - {\vec s}(i-1)$$ is the velocity at the moment $i$. Therefore,
\[
\begin{array}{lll}
{\vec s}(0) = 0 \\
\noalign{\medskip} {\vec s}(1) = {\vec v}_1 \\
\noalign{\medskip} {\vec s}(2) = {\vec v}_1 + {\vec v}_2 \\
\noalign{\medskip} \ldots \\
\noalign{\medskip} {\vec s}(m) = {\vec v}_1 + {\vec v}_2 + \ldots + {\vec v}_{m} = {\vec k}
\end{array}
\]
and the Schur polynomial is given by the sum over all such paths:
$$
P_{{\vec k}}\left\{ t_{\vec i} \right\} = \sum\limits_{m = 1}^{k_1 + k_2 + \ldots + k_n} \ \ \mathop{\sum\limits_{{\vec s}(i)}}_{{\vec s}(0) = 0, \ {\vec s}(m) = {\vec k}} \dfrac{t_{{\vec v_1}} t_{{\vec v_2}} \ldots t_{{\vec v_m}}}{m!}
$$
The length $m$ of a path is not smaller than 1, but not greater than $k_1 + k_2 + \ldots + k_n$. The correspondence between multi-partitions and $n$-dimensional lattice paths is illustrated by the following picture:
\begin{center}
\includegraphics[width=250pt]{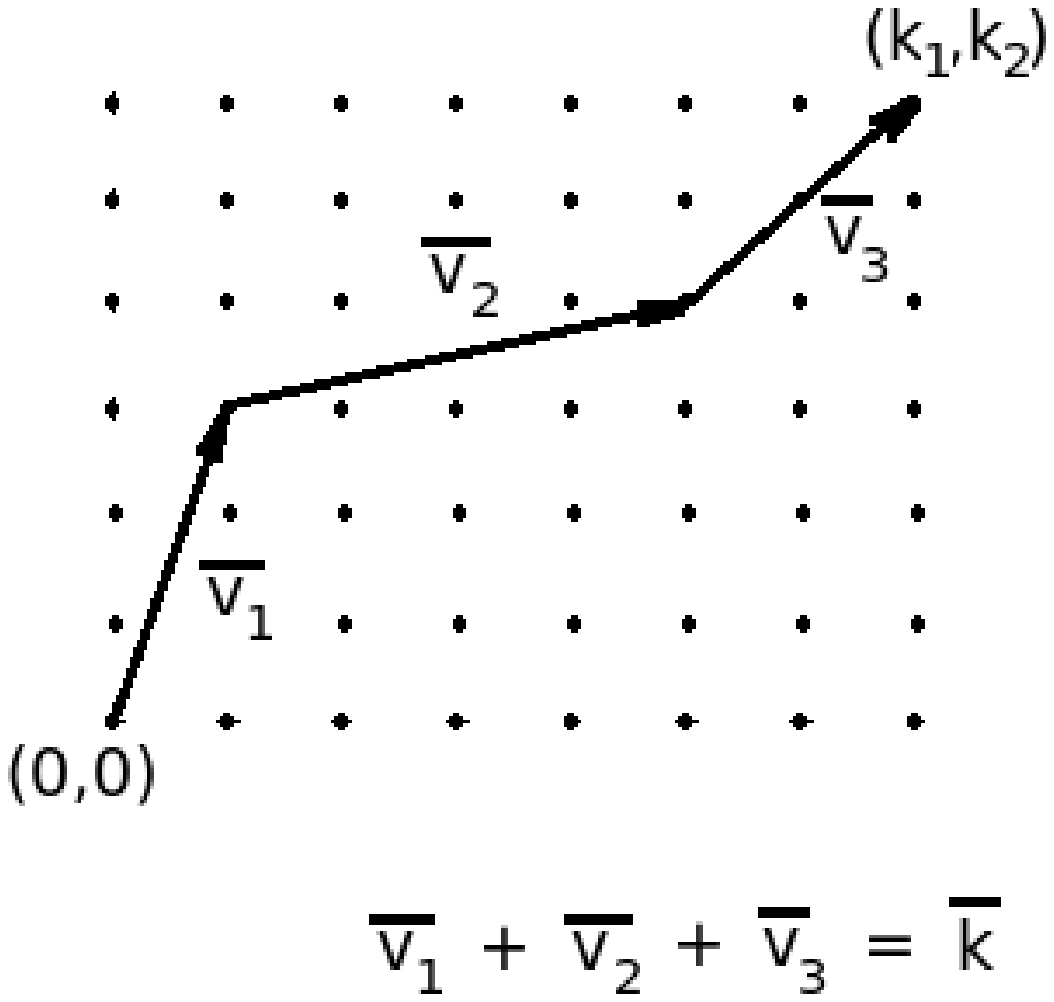}
\end{center}

Since resultant is a Schur polynomial, it is given by a sum over paths on a lattice. Such representation is a convenient way to visualise Schur polynomials (determinants, resultants). It provides a clear and intuitive picture for these quantities. Probably, it will be useful in future applications.

\section{Calculation of resultants }

Now, let us apply our method to calculate several resultants.

\paragraph{Calculation of $R_{1,1}$.} This is one of the simplest resultants: just a $2 \times 2$ determinant. We denote the polynomials as $f$ and $g$. The system of equations has the form
\[
\left\{ \begin{array}{c}
f(x_1, x_2) = f_{1} x_1 + f_{2} x_2 = 0 \\
\noalign{\medskip}g(x_1, x_2) = g_{1} x_1 + g_{2} x_2 = 0 \\
\end{array} \right.
\]
The traces are calculated, using differential operators
\[
\begin{array}{cc}
{\hat f} = f_1 \dfrac{ \partial }{\partial A_{11}} + f_2 \dfrac{ \partial }{\partial A_{12}}\\
\\
{\hat g} = g_1 \dfrac{ \partial }{\partial A_{21}} + g_2 \dfrac{ \partial }{\partial A_{22}}\\
\end{array}
\]
Traces of degree one:
\[
\left\{
\begin{array}{ll}
T_{1,0}(f) = {\hat f} \cdot {\rm tr} A = \left( f_1 \dfrac{ \partial }{\partial A_{11}} + f_2 \dfrac{ \partial }{\partial A_{12}} \right) \left( A_{11} + A_{22} \right) = f_{1} \\
\\
T_{0,1}(f) = {\hat g} \cdot {\rm tr} A = \left( g_1 \dfrac{ \partial }{\partial A_{11}} + g_2 \dfrac{ \partial }{\partial A_{12}} \right) \left( A_{11} + A_{22} \right) = g_{2} \\
\end{array}
\right.
\]
so that

$$ T_{1}(f) = T_{1,0} + T_{0,1} = f_{1} + g_{2} $$
Traces of degree two:
\[
\left\{
\begin{array}{ll}
T_{2,0}(f) = \dfrac{1}{4} {\hat f}^2 \cdot {\rm tr} A^2 = \dfrac{1}{4} \left( f_1 \dfrac{ \partial }{\partial A_{11}} + f_2 \dfrac{ \partial }{\partial A_{12}} \right)^2 \left( A_{11}^2 + 2 A_{12} A_{21} + A_{22}^2 \right) = f_{1}^2/2\\
\\
T_{1,1}(f) = \dfrac{1}{2} {\hat f} {\hat g} \cdot {\rm tr} A^2 = \dfrac{1}{2} \left( f_1 \dfrac{ \partial }{\partial A_{11}} + f_2 \dfrac{ \partial }{\partial A_{12}} \right) \left( g_1 \dfrac{ \partial }{\partial A_{21}} + g_2 \dfrac{ \partial }{\partial A_{22}} \right) \left( A_{11}^2 + 2 A_{12} A_{21} + A_{22}^2 \right) = f_{2} g_{1} \\
\\
T_{0,2}(f) = \dfrac{1}{4} {\hat g}^2 \cdot {\rm tr} A^2 = \dfrac{1}{4} \left( g_1 \dfrac{ \partial }{\partial A_{21}} + g_2 \dfrac{ \partial }{\partial A_{22}} \right)^2 \left( A_{11}^2 + 2 A_{12} A_{21} + A_{22}^2 \right) = g_{2}^2/2 \\
\end{array}
\right.
\]
so that

$$
T_{2}(f) = 2 \left( T_{2,0} + T_{1,1} + T_{0,2} \right) = f_{1}^2 + 2 f_{2} g_{1} + g_{2}^2
$$
Of course, since equations are linear, the traces are conventional quantities from linear algebra:
$$T_{1}(f) = {\rm tr} \left( \begin{array}{cc} f_{1} & f_{2} \\ g_{1} & g_{2} \\ \end{array} \right) \ \ \ {\rm and} \ \ \ T_{2}(f) = {\rm tr} \left( \begin{array}{cc} f_{1} & f_{2} \\ g_{1} & g_{2} \\ \end{array} \right)^2 = {\rm tr} \left( \begin{array}{cc} f_{1}^2 + f_{2} g_{1} & f_{1} f_{2} + f_{2} g_{2} \\ f_{1} g_{1} + g_{1} g_{2} & f_{2} g_{1} + g_{2}^2 \\ \end{array} \right)$$
The resultant has degree 2, and is expressed through traces as
$$ R_{1,1}\big\{f,g\big\} = P_{2} \left\{ \dfrac{- T_{k}(f)}{k} \right\} = T_{1}^2/2 - T_2/2$$
or directly through multigraded traces as
$$ R_{1,1}\big\{f,g\big\} = {\cal P}_{1,1} \left\{ - T_{k_1 k_2}(f) \right\} = T_{1,1} + T_{1,0} T_{0,1}$$
By substituting the expressions for traces, we obtain $ R_{1,1}\big\{f,g\big\} = f_{1} g_{2} - f_{2} g_{1} $.

\paragraph{Calculation of $R_{2,2}$.} We denote the polynomials as $f$ and $g$. The system of equations has the form
\[
\left\{ \begin{array}{c}
f(x_1, x_2) = f_{11} x_1^2 + f_{12} x_1 x_2 + f_{22} x_2^2 = 0 \\
\noalign{\medskip}g(x_1, x_2) = g_{11} x_1^2 + g_{12} x_1 x_2 + g_{22} x_2^2 = 0 \\
\end{array} \right.
\]
The traces are calculated, using differential operators
\[
\begin{array}{cc}
{\hat f} = f_{11} \left( \dfrac{ \partial }{\partial A_{11}} \right)^2 + f_{12} \dfrac{ \partial }{\partial A_{11}} \dfrac{ \partial }{\partial A_{12}} + f_{22} \left( \dfrac{ \partial }{\partial A_{12}} \right)^2 \\
\\
{\hat g} = f_{11} \left( \dfrac{ \partial }{\partial A_{21}} \right)^2 + f_{12} \dfrac{ \partial }{\partial A_{21}} \dfrac{ \partial }{\partial A_{22}} + f_{22} \left( \dfrac{ \partial }{\partial A_{22}} \right)^2\\
\end{array}
\]
acting on the quantities like
\[
\begin{array}{ccc}
{\rm tr} A^2 = A_{11}^2 + 2 A_{12} A_{21} + A_{22}^2\\
\\
\\
{\rm tr} A^4 = A_{1 1}^4+4 A_{1 1}^2 A_{1 2} A_{2 1}+2 A_{1 2}^2 A_{2 1}^2+4 A_{1 1} A_{1 2} A_{2 2} A_{2 1}+4 A_{1 2} A_{2 2}^2 A_{2 1}+A_{2 2}^4\\
\end{array}
\]
Traces of degree one:
\[
\left\{
\begin{array}{ll}
T_{1,0}(f) = {\hat f} \cdot {\rm tr} A^2 = 2 f_{11} \\
\\
T_{0,1}(f) = {\hat g} \cdot {\rm tr} A^2 = 2 g_{22} \\
\end{array}
\right.
\]
so that
$$T_{1}(f) = T_{1,0} + T_{0,1} = 2 f_{11} + 2 g_{22} $$
Traces of degree two
\[
\left\{
\begin{array}{ll}
T_{2,0}(f) = \dfrac{1}{24} {\hat f}^2 \cdot {\rm tr} A^4 = f_{1 1}^2\\
\\
T_{1,1}(f) = \dfrac{1}{4} {\hat f} {\hat g} \cdot {\rm tr} A^4 = 2 f_{2 2} g_{1 1}+f_{1 2} g_{1 2} \\
\\
T_{0,2}(f) = \dfrac{1}{24} {\hat g}^2 \cdot {\rm tr} A^4 = g_{2 2}^2 \\
\end{array}
\right.
\]
so that
$$ T_{2}(f) = 2 \left( T_{2,0} + T_{1,1} + T_{0,2} \right) = 2 f_{1 1}^2 + 4 f_{2 2} g_{1 1} + 2 f_{1 2} g_{1 2} + 2 g_{2}^2 $$
Traces of degree three
\[
\left\{
\begin{array}{ll}
T_{3, 0}(f) = \dfrac{1}{1080} {\hat f}^3 \cdot {\rm tr} A^6 = 2/3 f_{1 1}^3\\
\\
T_{2, 1}(f) = \dfrac{1}{72} {\hat f}^2 {\hat g} \cdot {\rm tr} A^6 = 2 f_{1 1} f_{2 2} g_{1 1}+f_{1 2}^2 g_{1 1}+f_{1 2} f_{1 1} g_{1 2} \\
\\
T_{1, 2}(f) = \dfrac{1}{72} {\hat f} {\hat g}^2 \cdot {\rm tr} A^6 = 2 f_{2 2} g_{1 1} g_{2 2}+f_{2 2} g_{1 2}^2+f_{1 2} g_{1 2} g_{2 2} \\
\\
T_{0, 3}(f) = \dfrac{1}{1080} {\hat g}^3 \cdot {\rm tr} A^6 = 2/3 g_{2 2}^3 \\
\end{array}
\right.
\]
so that
$$ T_{3}(f) = 3 \left( T_{3, 0} + T_{2, 1} + T_{1, 2} + T_{0, 3} \right) = 2 f_{1 1}^3+3 g_{1 2} f_{1 1} f_{1 2}+6 g_{1 1} f_{1 1} f_{2 2}+3 g_{1 1} f_{1 2}^2+3 g_{1 2} g_{2 2} f_{1 2}+3 g_{1 2}^2 f_{2 2}+6 g_{1 1} g_{2 2} f_{2 2}+2 g_{2 2}^3 $$
Traces of degree four
\[
\left\{
\begin{array}{ll}
T_{4,  0}(f) = \dfrac{1}{80640} {\hat f}^4 \cdot {\rm tr} A^8 = 1/2 f_{1 1}^4\\
\\
T_{3,  1}(f) = \dfrac{1}{2880} {\hat f}^3 {\hat g} \cdot {\rm tr} A^8 = f_{1 2} f_{1 1}^2 g_{1 2}+2 f_{1 2}^2 f_{1 1} g_{1 1}+2 f_{1 1}^2 f_{2 2} g_{1 1} \\
\\
T_{2,  2}(f) = \dfrac{1}{1152} {\hat f}^2 {\hat g}^2 \cdot {\rm tr} A^8 = f_{2 2}^2 g_{1 1}^2 + f_{1 1} f_{2 2} \left( g_{1 2}^2 + 2 g_{1 1} g_{2 2} \right) +1/2 f_{1 2}^2 \left( g_{1 2}^2 + 2 g_{1 1} g_{2 2} \right) +3 f_{1 2} f_{2 2} g_{1 2} g_{1 1}+f_{1 2} f_{1 1} g_{1 2} g_{2 2} \\
\\
T_{1,  3}(f) = \dfrac{1}{2880} {\hat f} {\hat g}^3  \cdot {\rm tr} A^8 = 2 f_{2 2} g_{1 1} g_{2 2}^2+2 f_{2 2} g_{1 2}^2 g_{2 2}+f_{1 2} g_{1 2} g_{2 2}^2 \\
\\
T_{0,  4}(f) = \dfrac{1}{80640} {\hat g}^4 \cdot {\rm tr} A^8 = 1/2 g_{2 2}^4 \\
\end{array}
\right.
\]
so that
\[
\begin{array}{ccc}
T_{4}(f) = 4 \left( T_{4, 0} + T_{3, 1} + T_{2, 2} + T_{1, 3} + T_{0, 4} \right) = 2 f_{1 1}^4+4 g_{1 2} f_{1 1}^2 f_{1 2}+8 g_{1 1} f_{1 1}^2 f_{2 2}+8 g_{1 1} f_{1 1} f_{1 2}^2+ \emph{} \\ \noalign{\medskip} 4 g_{1 2} g_{2 2} f_{1 1} f_{1 2}+4 g_{1 2}^2 f_{1 1} f_{2 2}+ 8 g_{1 1} g_{2 2} f_{1 1} f_{2 2}+2 g_{1 2}^2 f_{1 2}^2+ 4 g_{1 1} g_{2 2} f_{1 2}^2+12 g_{1 2} g_{1 1} f_{1 2} f_{2 2}+4 g_{1 1}^2 f_{2 2}^2+ \emph{} \\ \noalign{\medskip} 4 g_{1 2} g_{2 2}^2 f_{1 2}+8 g_{1 1} g_{2 2}^2 f_{2 2}+8 g_{1 2}^2 g_{2 2} f_{2 2}+2 g_{2 2}^4
\end{array}
\]
Note that, in contrast with linear algebra, $T_k(f)$ are not tensor invariants. Expression of $T_k(f)$ in terms of tensor contractions remains to be investigated. Resultant has degree 4, and is expressed through traces as
$$ R_{2,2}\big\{f,g\big\} = P_{4} \left\{ \dfrac{- T_{k}(f)}{k} \right\} = \dfrac{1}{24} \left( T_{1}^4 - 6 T_{1}^2 T_{2} + 8 T_{1} T_{3} + 3 T_{2}^2 - 6 T_{4} \right)$$
\pagebreak
or directly through multigraded traces as
$$ R_{2,2}\big\{f,g\big\} = {\cal P}_{2,2} \left\{ - T_{k_1 k_2}(f) \right\} = $$
$$= T_{2,2}+T_{0,2} T_{2,0}+T_{1,2} T_{1,0}+T_{2,1} T_{0,1}+ \dfrac{1}{2} T_{1,1}^2+\dfrac{1}{2} T_{1,0}^2 T_{0,2}+\dfrac{1}{2} T_{0,1}^2 T_{2,0}+T_{1,1} T_{0,1} T_{1,0}+\dfrac{1}{4} T_{0,1}^2 T_{1,0}^2 $$
By substituting the expressions for traces, we obtain
$$ R_{2,2}\big\{f,g\big\} = f_{1 1}^2 g_{2 2}^2 - f_{1 1} f_{1 2} g_{1 2} g_{2 2} + f_{1 1} f_{2 2} g_{1 2}^2 - 2 f_{1 1} f_{2 2} g_{1 1} g_{2 2} + f_{1 2}^2 g_{1 1} g_{2 2} - f_{1 2} f_{2 2} g_{1 2} g_{1 1} + f_{2 2}^2 g_{1 1}^2 $$

\paragraph{Calculation of $R_{3,3}$.} We denote the polynomials as $f$ and $g$. The system of equations has the form
\[
\left\{ \begin{array}{c}
f(x_1, x_2) = f_{111} x_1^3 + f_{112} x_1^2 x_2 + f_{122} x_1 x_2^2 + f_{222} x_2^3 = 0 \\
\noalign{\medskip}g(x_1, x_2) = g_{111} x_1^3 + g_{112} x_1^2 x_2 + g_{122} x_1 x_2^2 + g_{222} x_2^3 = 0 \\
\end{array} \right.
\]
The traces are calculated with (\ref{PP2}), using differential operators
\[
\begin{array}{cc}
{\hat f} = f_{111} \left( \dfrac{ \partial }{\partial A_{11}} \right)^3 + f_{112} \left( \dfrac{ \partial }{\partial A_{11}} \right)^2 \left( \dfrac{ \partial }{\partial A_{12}} \right) + f_{122} \left( \dfrac{ \partial }{\partial A_{11}} \right) \left( \dfrac{ \partial }{\partial A_{12}} \right)^2 + f_{222} \left( \dfrac{ \partial }{\partial A_{12}} \right)^3  \\
\\
{\hat g} = g_{111} \left( \dfrac{ \partial }{\partial A_{21}} \right)^3 + g_{112} \left( \dfrac{ \partial }{\partial A_{21}} \right)^2 \left( \dfrac{ \partial }{\partial A_{22}} \right) + g_{122} \left( \dfrac{ \partial }{\partial A_{21}} \right) \left( \dfrac{ \partial }{\partial A_{22}} \right)^2 + g_{222} \left( \dfrac{ \partial }{\partial A_{22}} \right)^3\\
\end{array}
\]
A few first traces are
\begin{align*}
& T_1 = 3 f_{1 1 1}+3 g_{2 2 2} \\
& \\
& T_2 = 3 f_{1 1 1}^2+6 g_{1 1 1} f_{2 2 2}+4 g_{1 1 2} f_{1 2 2}+2 g_{1 2 2} f_{1 1 2}+3 g_{2 2 2}^2 \\
& \\
& T_3 = 3 f_{1 1 1}^3+9 g_{1 1 1} f_{2 2 2} f_{1 1 1}+9 g_{1 1 1} f_{1 2 2} f_{1 1 2}+6 g_{1 1 2} f_{1 2 2} f_{1 1 1}+3 g_{1 1 2} f_{1 1 2}^2+3 g_{1 2 2} f_{1 1 2} f_{1 1 1}+9 g_{1 1 2} g_{1 2 2} f_{2 2 2}+ \\ & 9 g_{1 1 1} g_{2 2 2} f_{2 2 2}+6 g_{1 1 2} g_{2 2 2} f_{1 2 2}+3 g_{1 2 2}^2 f_{1 2 2}+3 g_{1 2 2} g_{2 2 2} f_{1 1 2}+3 g_{2 2 2}^3 \\
& \\
& T_4 = 3 f_{1 1 1}^4+12 g_{1 1 1} f_{2 2 2} f_{1 1 1}^2+24 g_{1 1 1} f_{1 2 2} f_{1 1 2} f_{1 1 1}+8 g_{1 1 2} f_{1 2 2} f_{1 1 1}^2+4 g_{1 1 1} f_{1 1 2}^3+8 g_{1 1 2} f_{1 1 2}^2 f_{1 1 1}+4 g_{1 2 2} f_{1 1 2} f_{1 1 1}^2+ \\ & 6 g_{1 1 1}^2 f_{2 2 2}^2+20 g_{1 1 2} g_{1 1 1} f_{2 2 2} f_{1 2 2}+ 16 g_{1 1 1} g_{1 2 2} f_{2 2 2} f_{1 1 2}+8 g_{1 1 2}^2 f_{2 2 2} f_{1 1 2}+12 g_{1 1 2} g_{1 2 2} f_{2 2 2} f_{1 1 1}+ 12 g_{1 1 1} g_{2 2 2} f_{2 2 2} f_{1 1 1} + \\ & 8 g_{1 1 1} g_{1 2 2} f_{1 2 2}^2+4 g_{1 1 2}^2 f_{1 2 2}^2+12 g_{1 1 2} g_{1 2 2} f_{1 2 2} f_{1 1 2}+12 g_{1 1 1} g_{2 2 2} f_{1 2 2} f_{1 1 2}+4 g_{1 2 2}^2 f_{1 2 2} f_{1 1 1}+8 g_{1 1 2} g_{2 2 2} f_{1 2 2} f_{1 1 1} + \\ & 4 g_{1 1 2} g_{2 2 2} f_{1 1 2}^2+2 g_{1 2 2}^2 f_{1 1 2}^2+4 g_{1 2 2} g_{2 2 2} f_{1 1 2} f_{1 1 1}+ 12 g_{1 1 1} g_{2 2 2}^2 f_{2 2 2}+24 g_{1 1 2} g_{1 2 2} g_{2 2 2} f_{2 2 2}+4 g_{1 2 2}^3 f_{2 2 2}+ \\ & 8 g_{1 2 2}^2 g_{2 2 2} f_{1 2 2}+8 g_{1 1 2} g_{2 2 2}^2 f_{1 2 2}+4 g_{1 2 2} g_{2 2 2}^2 f_{1 1 2}+3 g_{2 2 2}^4 \\
\end{align*}
and so on. The resultant has degree 6, and is expressed through traces as
$$ R_{3,3}\big\{f,g\big\} = P_{6} \left\{ \dfrac{- T_{k}(f)}{k} \right\} = $$
$$\dfrac{1}{720} \left( T_{1}^6-15 T_{1}^4 T_{2}+40 T_{1}^3 T_{3}+45 T_{1}^2 T_{2}^2-90 T_{1}^2 T_{4}-120 T_{1} T_{2} T_{3}-15 T_{2}^3+144 T_{1} T_{5}+90 T_{2} T_{4}+40 T_{3}^2-120 T_{6} \right)$$
By substituting the expressions for traces, we obtain
\begin{align*}
& R_{3,3}\big\{f,g\big\} = f_{1 1 1}^3 g_{2 2 2}^3-f_{1 1 1}^2 f_{1 1 2} g_{2 2 2}^2 g_{1 2 2}+f_{1 1 1} f_{1 1 2}^2 g_{2 2 2}^2 g_{1 1 2}-2 f_{1 1 1}^2 f_{1 2 2} g_{2 2 2}^2 g_{1 1 2} + 3 f_{1 1 1} f_{1 1 2} f_{1 2 2} g_{2 2 2}^2 g_{1 1 1} - \emph{} \\ & 3 f_{1 1 1}^2 f_{2 2 2} g_{2 2 2}^2 g_{1 1 1} - f_{1 1 2}^3 g_{2 2 2}^2 g_{1 1 1}+f_{1 1 1}^2 f_{1 2 2} g_{2 2 2} g_{1 2 2}^2+3 f_{1 1 1}^2 f_{2 2 2} g_{2 2 2} g_{1 2 2} g_{1 1 2}-f_{1 1 1} f_{1 1 2} f_{1 2 2} g_{2 2 2} g_{1 2 2} g_{1 1 2}- \emph{} \\ & 2 f_{1 1 1} f_{1 2 2}^2 g_{2 2 2} g_{1 2 2} g_{1 1 1}-f_{1 1 1} f_{1 1 2} f_{2 2 2} g_{2 2 2} g_{1 2 2} g_{1 1 1}+f_{1 1 2}^2 f_{1 2 2} g_{2 2 2} g_{1 2 2} g_{1 1 1}-2 f_{1 1 1} f_{1 1 2} f_{2 2 2} g_{2 2 2} g_{1 1 2}^2+\emph{} \\ &f_{1 1 1} f_{1 2 2}^2 g_{2 2 2} g_{1 1 2}^2+f_{1 1 1} f_{1 2 2} f_{2 2 2} g_{2 2 2} g_{1 1 2} g_{1 1 1}+2 f_{1 1 2}^2 f_{2 2 2} g_{2 2 2} g_{1 1 2} g_{1 1 1}-f_{1 1 2} f_{1 2 2}^2 g_{2 2 2} g_{1 1 2} g_{1 1 1}+f_{1 2 2}^3 g_{2 2 2} g_{1 1 1}^2 - \emph{} \\ & 3 f_{1 1 2} f_{1 2 2} f_{2 2 2} g_{2 2 2} g_{1 1 1}^2+3 f_{1 1 1} f_{2 2 2}^2 g_{2 2 2} g_{1 1 1}^2-f_{1 1 1}^2 f_{2 2 2} g_{1 2 2}^3+f_{1 1 1} f_{1 1 2} f_{2 2 2} g_{1 2 2}^2 g_{1 1 2}-f_{1 1 2}^2 f_{2 2 2} g_{1 2 2}^2 g_{1 1 1}+ \emph{} \\ & 2 f_{1 1 1} f_{1 2 2} f_{2 2 2} g_{1 2 2}^2 g_{1 1 1}-f_{1 1 1} f_{1 2 2} f_{2 2 2} g_{1 2 2} g_{1 1 2}^2+f_{1 1 2} f_{1 2 2} f_{2 2 2} g_{1 2 2} g_{1 1 2} g_{1 1 1}-3 f_{1 1 1} f_{2 2 2}^2 g_{1 2 2} g_{1 1 2} g_{1 1 1}-\emph{} \\ & f_{1 2 2}^2 f_{2 2 2} g_{1 2 2} g_{1 1 1}^2+2 f_{1 1 2} f_{2 2 2}^2 g_{1 2 2} g_{1 1 1}^2+f_{1 1 1} f_{2 2 2}^2 g_{1 1 2}^3-f_{1 1 2} f_{2 2 2}^2 g_{1 1 2}^2 g_{1 1 1}+f_{1 2 2} f_{2 2 2}^2 g_{1 1 2} g_{1 1 1}^2-f_{2 2 2}^3 g_{1 1 1}^3
\end{align*}

\paragraph{Calculation of $R_{2,2,2}$.} This is already an example of multidimensional resultants. We denote the polynomials as $f$, $g$ and $h$. The system of equations has the form
\[
\left\{ \begin{array}{c}
f(x_1, x_2, x_3) = f_{11} x_1^2 + f_{12} x_1 x_2 + f_{13} x_1 x_3 + f_{22} x_2^2 + f_{23} x_2 x_3 + f_{33} x_3^2 = 0 \\
\noalign{\medskip}g(x_1, x_2, x_3) = g_{11} x_1^2 + g_{12} x_1 x_2 + g_{13} x_1 x_3 + g_{22} x_2^2 + g_{23} x_2 x_3 + g_{33} x_3^2 = 0 \\
\noalign{\medskip}h(x_1, x_2, x_3) = h_{11} x_1^2 + h_{12} x_1 x_2 + h_{13} x_1 x_3 + h_{22} x_2^2 + h_{23} x_2 x_3 + h_{33} x_3^2 = 0 \\
\end{array} \right.
\]
The traces are calculated with (\ref{PP2}), using differential operators
\[
\begin{array}{cc}
{\hat f} = f_{11} \left( \dfrac{ \partial }{\partial A_{11}} \right)^2 + f_{12} \dfrac{ \partial^2 }{\partial A_{11} \partial A_{12}} + f_{13} \dfrac{ \partial^2 }{\partial A_{11} \partial A_{13}} + f_{22} \left( \dfrac{ \partial }{\partial A_{12}} \right)^2 + f_{23} \dfrac{ \partial^2 }{\partial A_{12} \partial A_{13}} + f_{33} \left( \dfrac{ \partial }{\partial A_{13}} \right)^2  \\
\\
{\hat g} = g_{11} \left( \dfrac{ \partial }{\partial A_{21}} \right)^2 + g_{12} \dfrac{ \partial^2 }{\partial A_{21} \partial A_{22}} + g_{13} \dfrac{ \partial^2 }{\partial A_{21} \partial A_{23}} + g_{22} \left( \dfrac{ \partial }{\partial A_{22}} \right)^2 + g_{23} \dfrac{ \partial^2 }{\partial A_{22} \partial A_{23}} + g_{33} \left( \dfrac{ \partial }{\partial A_{23}} \right)^2 \\
\\
{\hat h} = h_{11} \left( \dfrac{ \partial }{\partial A_{31}} \right)^2 + h_{12} \dfrac{ \partial^2 }{\partial A_{31} \partial A_{32}} + h_{13} \dfrac{ \partial^2 }{\partial A_{31} \partial A_{33}} + h_{22} \left( \dfrac{ \partial }{\partial A_{32}} \right)^2 + h_{23} \dfrac{ \partial^2 }{\partial A_{32} \partial A_{33}} + h_{33} \left( \dfrac{ \partial }{\partial A_{33}} \right)^2
\end{array}
\]
A few first traces are
\begin{align*}
& T_1 = 4 f_{1 1}+4 g_{2 2}+4 h_{3 3} \\
& \\
& T_2 = 4 f_{1 1}^2+4 g_{1 2} f_{1 2}+4 h_{1 3} f_{1 3}+8 g_{1 1} f_{2 2}+8 h_{1 1} f_{3 3}+4 g_{2 2}^2+8 g_{3 3} h_{2 2}+4 g_{2 3} h_{2 3}+4 h_{3 3}^2 \\
& \\
& T_3 = 4 f_{1 1}^3+6 g_{1 2} f_{1 1} f_{1 2}+6 h_{1 3} f_{1 1} f_{1 3}+12 g_{1 1} f_{1 1} f_{2 2}+12 h_{1 1} f_{1 1} f_{3 3}+6 g_{1 1} f_{1 2}^2+6 h_{1 1} f_{1 3}^2+3 g_{2 3} h_{1 3} f_{1 2}+ \emph{} \\ & 6 g_{3 3} h_{1 2} f_{1 2}+  6 g_{1 2} g_{2 2} f_{1 2}+3 g_{1 3} h_{2 3} f_{1 2}+3 g_{2 3} h_{1 2} f_{1 3}+6 h_{1 3} h_{3 3} f_{1 3}+6 g_{1 3} h_{2 2} f_{1 3}+3 g_{1 2} h_{2 3} f_{1 3}+ 6 g_{1 2}^2 f_{2 2}+ \emph{} \\ & 12 g_{1 1} g_{2 2} f_{2 2}+6 g_{1 3} h_{1 3} f_{2 2}+12 g_{3 3} h_{1 1} f_{2 2}+6 g_{2 3} h_{1 1} f_{2 3}+3 g_{1 2} h_{1 3} f_{2 3}+ 9 g_{1 3} h_{1 2} f_{2 3}+6 g_{1 1} h_{2 3} f_{2 3}+6 h_{1 3}^2 f_{3 3}+\emph{} \\ & 6 g_{1 2} h_{1 2} f_{3 3}+12 g_{1 1} h_{2 2} f_{3 3}+12 h_{1 1} h_{3 3} f_{3 3}+12 g_{2 2} g_{3 3} h_{2 2}+4 g_{2 2}^3+6 g_{2 3} h_{2 3} h_{3 3}+6 g_{2 2} g_{2 3} h_{2 3}+\emph{} \\ & 6 g_{2 3}^2 h_{2 2}+6 g_{3 3} h_{2 3}^2+12 g_{3 3} h_{2 2} h_{3 3} +4 h_{3 3}^3
\end{align*}
and so on. The resultant has degree 12, and is expressed through traces as
$$ R_{2,2,2}\big\{f,g,h\big\} = P_{12} \left\{ \dfrac{- T_{k}(f)}{k} \right\} = $$
\begin{center}
$
-\dfrac{1}{12} T_{12}+\dfrac{1}{72} T_{6}^2+\dfrac{1}{35} T_{5} T_{7}+\dfrac{1}{32} T_{4} T_{8}+\dfrac{1}{27} T_{3} T_{9}+\dfrac{1}{20} T_{2} T_{10}+\dfrac{1}{11} T_{1} T_{11}-\dfrac{1}{384} T_{4}^3-\dfrac{1}{60} T_{3} T_{4} T_{5}-\dfrac{1}{108} T_{3}^2 T_{6}-\dfrac{1}{100} T_{2} T_{5}^2-\dfrac{1}{48} T_{2} T_{4} T_{6}-\dfrac{1}{42} T_{2} T_{3} T_{7}-\dfrac{1}{64} T_{2}^2 T_{8}-\dfrac{1}{30} T_{1} T_{5} T_{6}-\dfrac{1}{28} T_{1} T_{4} T_{7}-\dfrac{1}{24} T_{1} T_{3} T_{8}-\dfrac{1}{18} T_{1} T_{2} T_{9}-\dfrac{1}{20} T_{1}^2 T_{10}+\dfrac{1}{1944} T_{3}^4+\dfrac{1}{144} T_{2} T_{3}^2 T_{4}+\dfrac{1}{256} T_{2}^2 T_{4}^2+\dfrac{1}{120} T_{2}^2 T_{3} T_{5}+\dfrac{1}{288} T_{2}^3 T_{6}+\dfrac{1}{96} T_{1} T_{3} T_{4}^2+\dfrac{1}{90} T_{1} T_{3}^2 T_{5}+\dfrac{1}{40} T_{1} T_{2} T_{4} T_{5}+\dfrac{1}{36} T_{1} T_{2} T_{3} T_{6}+\dfrac{1}{56} T_{1} T_{2}^2 T_{7}+\dfrac{1}{100} T_{1}^2 T_{5}^2+\dfrac{1}{48} T_{1}^2 T_{4} T_{6}+\dfrac{1}{42} T_{1}^2 T_{3} T_{7}+\dfrac{1}{32} T_{1}^2 T_{2} T_{8}+\dfrac{1}{54} T_{1}^3 T_{9}-\dfrac{1}{864} T_{2}^3 T_{3}^2-\dfrac{1}{1536} T_{2}^4 T_{4}-\dfrac{1}{324} T_{1} T_{2} T_{3}^3-\dfrac{1}{96} T_{1} T_{2}^2 T_{3} T_{4}-\dfrac{1}{240} T_{1} T_{2}^3 T_{5}-\dfrac{1}{144} T_{1}^2 T_{3}^2 T_{4}-\dfrac{1}{128} T_{1}^2 T_{2} T_{4}^2-\dfrac{1}{60} T_{1}^2 T_{2} T_{3} T_{5}-\dfrac{1}{96} T_{1}^2 T_{2}^2 T_{6}-\dfrac{1}{120} T_{1}^3 T_{4} T_{5}-\dfrac{1}{108} T_{1}^3 T_{3} T_{6}-\dfrac{1}{84} T_{1}^3 T_{2} T_{7}-\dfrac{1}{192} T_{1}^4 T_{8}+\dfrac{1}{46080} T_{2}^6+\dfrac{1}{1152} T_{1} T_{2}^4 T_{3}+\dfrac{1}{288} T_{1}^2 T_{2}^2 T_{3}^2+\dfrac{1}{384} T_{1}^2 T_{2}^3 T_{4}+\dfrac{1}{972} T_{1}^3 T_{3}^3+\dfrac{1}{144} T_{1}^3 T_{2} T_{3} T_{4}+\dfrac{1}{240} T_{1}^3 T_{2}^2 T_{5}+\dfrac{1}{768} T_{1}^4 T_{4}^2+\dfrac{1}{360} T_{1}^4 T_{3} T_{5}+\dfrac{1}{288} T_{1}^4 T_{2} T_{6}+\dfrac{1}{840} T_{1}^5 T_{7}-\dfrac{1}{7680} T_{1}^2 T_{2}^5-\dfrac{1}{864} T_{1}^3 T_{2}^3 T_{3}-\dfrac{1}{864} T_{1}^4 T_{2} T_{3}^2-\dfrac{1}{768} T_{1}^4 T_{2}^2 T_{4}-\dfrac{1}{1440} T_{1}^5 T_{3} T_{4}-\dfrac{1}{1200} T_{1}^5 T_{2} T_{5}-\dfrac{1}{4320} T_{1}^6 T_{6}+\dfrac{1}{9216} T_{1}^4 T_{2}^4+\dfrac{1}{2880} T_{1}^5 T_{2}^2 T_{3}+\dfrac{1}{12960} T_{1}^6 T_{3}^2+\dfrac{1}{5760} T_{1}^6 T_{2} T_{4}+\dfrac{1}{25200} T_{1}^7 T_{5}-\dfrac{1}{34560} T_{1}^6 T_{2}^3-\dfrac{1}{30240} T_{1}^7 T_{2} T_{3}-\dfrac{1}{161280} T_{1}^8 T_{4}+\dfrac{1}{322560} T_{1}^8 T_{2}^2+\dfrac{1}{1088640} T_{1}^9 T_{3}-\dfrac{1}{7257600} T_{1}^{10} T_{2}+\dfrac{1}{479001600} T_{1}^{12}
$
\end{center}

Despite numeric coefficients become large, this is a rather compact formula for $R_{2,2,2}$. By substituting the expressions for traces, we obtain a polynomial expression in $f$, $g$ and $h$, which contains 21894 terms and takes approximately 50 pages of A4 paper. Of course, we cannot write it here explicitly. See the appendix of the book version of \cite{NOLINAL}, where such a quantity is presented for curiosity.

\section{Acknowledgements}

It is a pleasure to thank V.Dolotin and all the participants of ITEP seminars on non-linear algebra for useful discussions. This work is partly supported by Russian Federal Atomic Energy Agency and Russian Academy of Sciences, by the joint grant 06-01-92059-CE, by NWO project 047.011.2004.026, by INTAS grant 05-1000008-7865, by ANR-05-BLAN-0029-01 project, by RFBR grant 07-02-00645 and by the Russian President's Grant of Support for the Scientific Schools NSh-3035.2008.2. The work of Sh.Shakirov is also partly supported by the Dynasty Foundation.

\end{document}